\documentstyle[aps]{revtex}
\begin{document}
\draft
\bibliographystyle{prsty}
\title{Evolution of speckle during spinodal decomposition}
\author{Gregory Brown,$^{1,2}$ Per Arne Rikvold,$^1$ 
        Mark Sutton,$^2$ and  Martin Grant$^2$
}
\address{
$^1$Supercomputer Computations Research Institute,\\
    Center for Materials Research and Technology,
    and Department of Physics\\
    Florida State University, Tallahassee, Florida 32306-4052\\
$^2$Centre for the Physics of Materials, McGill University,\\
    3600 rue University, Montr{\'{e}}al, Qu{\'{e}}bec, Canada H3A 2T8\\
}

\date{\today}
\maketitle
\begin{abstract}

Time-dependent properties of the speckled intensity patterns created
by scattering coherent radiation from materials undergoing spinodal
decomposition are investigated by numerical integration of the
Cahn-Hilliard-Cook equation. For binary systems which obey a local
conservation law, the characteristic domain size is known to grow in
time $\tau$ as $R = [B \tau]^n$ with $n=1/3$, where $B$ is a constant.
The intensities of individual speckles are found to be nonstationary,
persistent time series. The two-time intensity covariance at wave
vector ${\bf k}$ can be collapsed onto a scaling function ${\rm
Cov}(\delta t,\bar{t}\,)$, where $\delta t = k^{1/n} B
|\tau_2-\tau_1|$ and $\bar{t} = k^{1/n} B (\tau_1+\tau_2)/2$. Both
analytically and numerically, the covariance is found to depend on
$\delta t$ only through $\delta t/\bar{t}$ in the small-$\bar{t}$
limit and $\delta t/\bar{t}\,^{1-n}$ in the large-$\bar{t}$ limit,
consistent with a simple theory of moving interfaces that applies to
any universality class described by a scalar order parameter. The
speckle-intensity covariance is numerically demonstrated to be equal
to the square of the two-time structure factor of the scattering
material, for which an analytic scaling function is obtained for large
$\bar{t}.$ In addition, the two-time, two-point order-parameter
correlation function is found to scale as
$C\left(r/\left(B^n\sqrt{\tau_1^{2n}+\tau_2^{2n}}\right),\tau_1/\tau_2\right)$,
even for quite large distances $r$. The asymptotic power-law exponent
for the autocorrelation function is found to be $\lambda \approx
4.47$, violating an upper bound conjectured by Fisher and Huse.

\end{abstract}


\section{Introduction}

One of the most common ways to obtain information about the spatial
structure of a material involves the scattering of radiation, such as
electrons, neutrons, or photons. The scattering is caused by
inhomogeneities in the scattering cross section of the
material. Depending on the nature of the incident radiation, this
scattering cross section is given by the local value of some property
of the material, such as its density or magnetic polarization. In the
first Born approximation the amplitude of the scattered wave is given
by the Fourier transform of that property, and the wave vector ${\bf
k}$ is proportional to the momentum transfer as a coherent wave of the
probe radiation is scattered. For conventional radiation sources, the
wave is only coherent over a small volume and different parts of the
sample scatter independently of each other. As a result, only the
ensemble average of the scattering intensity can be observed.

\begin{figure}[tb]
\vskip 6in
\includegraphics{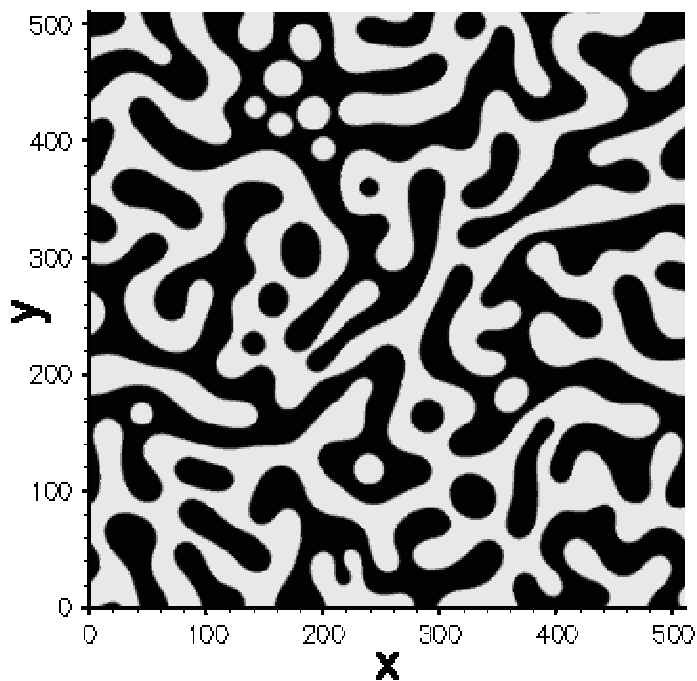}
\bigskip
\bigskip
\caption[]{Snapshot of a system undergoing spinodal decomposition at the
maximum simulation time, $\tau=4000$. All simulations reported here
were conducted on $512 \times 512$ lattices.}
\label{fig:config}
\end{figure}

\begin{figure}[tb]
\vskip 6in
\includegraphics{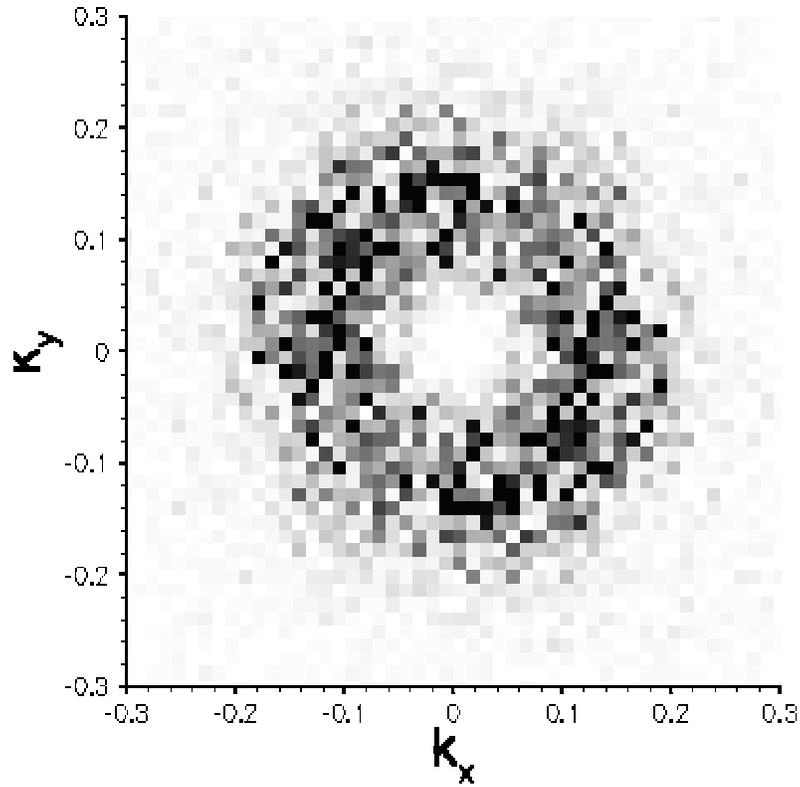}
\bigskip
\bigskip
\caption[]{Central region of the scattering intensity for one
realization of the simulation at time $\tau=4000$. The speckling of
the scattering pattern is apparent, with darker shades indicating brighter
speckles.}
\label{fig:speckle}
\end{figure}

On the other hand, when the coherence volume of the wave is
sufficiently large to encompass the whole scattering volume, the
unaveraged Fourier transform of the specific geometric structure of
the material is observed. If the inhomogeneous scattering property of
a particular sample at position ${\bf r}$ and time $\tau$ is
represented as the scalar field $\psi({\bf r},\tau)$, and its Fourier
transform at wave vector ${\bf k}$ as $\hat\psi({\bf k},\tau)$, then
the observed scattering intensity is
\begin{equation}
\label{eq:intensity}
I({\bf k},\tau)=\left|\hat\psi({\bf k},\tau)\right|^2 \;,
\end{equation}
where proportionality constants have been ignored. The domain pattern
of the order parameter $\psi({\bf r},\tau)$ from one of our
simulations of spinodal decomposition in a two-phase material is shown
in Fig.~\ref{fig:config}, with the shading differences corresponding
to differences in the local scattering cross sections.
Figure~\ref{fig:speckle} shows the central section of the scattering
intensity corresponding to Fig.~\ref{fig:config}, with darker shades
indicating higher intensities. The speckled appearance of the
scattering intensity is a characteristic feature in the scattering of
coherent radiation.

Since the speckle pattern depends on the specific configuration of the
material, it changes as the domain structure evolves over time. The
result is that the individual speckles fluctuate around their
time-dependent averages. The intensity fluctuations are almost
uncorrelated in ${\bf k}$-space, but the evolution of the speckle
pattern gives rise to nontrivial two-time correlations in the
scattering intensity at individual wave vectors.

\begin{figure}[tb]
\vskip 6in
\includegraphics{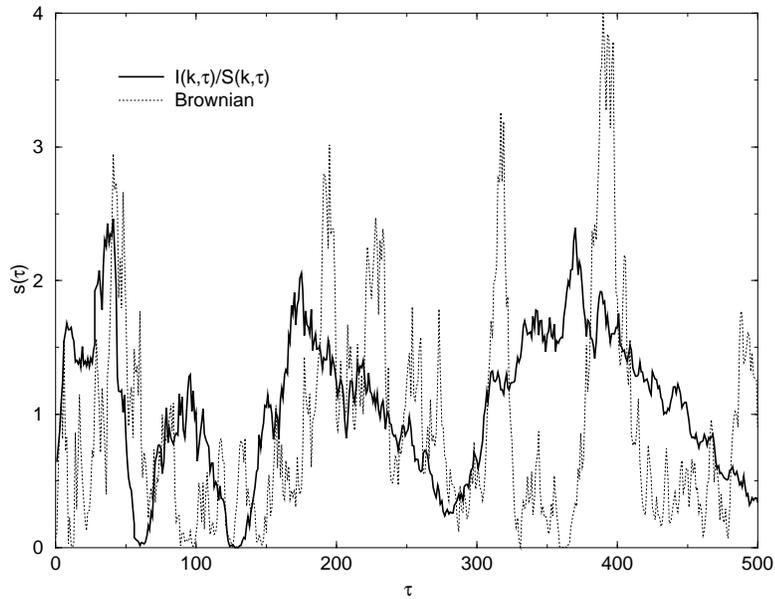}
\bigskip
\bigskip
\caption[]{Time series of the scattering intensity at one wave vector
${\bf k}$ for one quench to zero temperature. The intensity has been
normalized by the time-dependent structure factor averaged over 100
quenches. The dotted line is a synthetic ``Brownian'' function
constructed to have the same exponential single-time probability
density as the scattering intensity and an exponential two-time
covariance with a characteristic time corresponding to that of the
simulation intensity. The persistence of the scattering-intensity time
series is quite apparent when compared to the synthetic Brownian time
series.}
\label{fig:persistence}
\end{figure}

The fluctuation of a single speckle around the average intensity is
shown in Fig.~\ref{fig:persistence}. It has been normalized at each
time by the corresponding average intensity. Also shown in
Fig.~\ref{fig:persistence} are Brownian noise fluctuations, whose
two-time covariance is exponential.  That noise is constructed to have
the same single-time probability density and characteristic time scale
as the speckle.  There is a clear qualitative difference in time
correlations between the two time series. The changes in the Brownian
function are relatively large and the fluctuations shorter lived than
those of the speckle intensities produced by the phase-ordering
simulation. This highlights a property of the speckle intensities
called persistence \cite{FEDE88}. While the changes in a Brownian time
series are independent, the changes in a persistent time series are
correlated with each other.

In the past, photon correlation experiments using laser light at
wavelengths ranging from the infrared to the ultraviolet have used the
two-time correlations of individual speckles to study fluctuations in
many materials \cite{CHU,MAND95}. New high-brilliance synchrotron photon
sources have made coherent x-ray experiments feasible. The comparative
advantages of x~rays over lasers for many systems include penetration
of optically opaque materials and the ability to investigate shorter
length scales. Coherent x-ray intensity fluctuation spectroscopy
(XIFS) experiments have already been performed on a number of
materials \cite{DIER95,CHU95,BRAU95,MOCH97}. Recently, the two-time
correlations in speckle intensity were measured by XIFS for a sodium
borosilicate glass undergoing spinodal decomposition \cite{MALI98}.

Our intent with this paper is to make a theoretical study of how
speckle experiments can be exploited to investigate the relaxation of
homogeneous materials towards a new, heterogeneous equilibrium after a
rapid change in a thermodynamic parameter such as
temperature. Previously \cite{BROW97,BROW97A} this was done for
materials subject to phase ordering, in which case $\psi({\bf r},t)$
is not constrained by any conservation laws (often called model~A
\cite{GUNT83}). Here we analyze scattering from materials subject to a
local conservation constraint (model~B \cite{GUNT83}), such as the
process of phase separation in a model binary mixture or alloy. The
dynamics of the phase separation that occurs after a high-temperature
homogeneous mixture is suddenly brought into the low-temperature
regime of two-phase coexistence is a fundamental
statistical-mechanical process that has important consequences for
multi-phase materials. Two examples are metallic alloys \cite{GUNT83}
and Vycor glass \cite{TOMO86}.

The scaling properties of correlations in the model system and their
relationship to speckle correlations are discussed in
Sec.~\ref{sec:scaling}.  Although our focus is on model~B, many of our
results are sufficiently general that the also apply to other
universality classes characterized by a scalar order parameter, such
as model~A.  Details of the numerical simulations are given in
Sec.~\ref{sec:numerics}. In Sec.~\ref{sec:results} the numerical
results are presented and compared with the theoretical scaling
relations from Sec.~\ref{sec:scaling}.  Our conclusions are summarized
in Sec.~\ref{sec:conclusion}.

\section{Scaling Theory}
\label{sec:scaling}

When the volume fractions of the two components are equal in a binary
mixture, the homogeneous high-temperature phase is unstable to
long-wavelength fluctuations after the quench. As a consequence, an
interpenetrating pattern of domains rich in one or the other of the
two components develops immediately after the quench.  (See Fig. 1.)
This intertwined structure distinguishes spinodal decomposition from
many other phase-separation processes. At later times, the domains
grow as the system evolves towards an equilibrium state with the
smallest possible total interfacial area compatible with local
conservation of the volume fractions of the constituent phases. The
patterns at different times are statistically similar, except for the
characteristic length scale of the domains.  The description of the
late-time regime of phase separation is greatly simplified by this
dynamic scaling behavior.

\subsection{Scaling of one-time functions}
\label{sec:one-time}

Usually, the average size and shape of the growing domains can be
measured using the structure factor, which is proportional to the
average intensity measured in scattering experiments. For isotropic
materials the structure factor depends only on the magnitude of the
wave vector,
\begin{equation}
\label{eq:skfun}
S(k,\tau)= \langle I({\bf k},\tau) \rangle \;,
\end{equation}
where the angular brackets represent the ensemble average.  For
locally conserved dynamics, the structure factor vanishes at both high
and low wave vectors. The intervening maximum occurs at $k_{\rm
max}(\tau)$, which is proportional to the inverse of the
characteristic domain size. Since the domains grow, the ring of
brightest scattering contracts as the system evolves toward
equilibrium.

In the scaling regime, the growth of this characteristic length obeys
a power law, $R(\tau)=[B \tau]^n$, where $B$ is a constant.  Such
power-law growth is found in many different phase-ordering processes,
and the exponent is usually independent of the specific properties of
the material.  Instead it depends on general features of the phase
ordering, such as the presence or absence of local conservation.  This
dynamical universality picture has been found to be widely applicable,
and a group of diverse processes that show common scaling behavior are
termed a universality class \cite{GUNT83}.  For situations described
by a scalar order parameter, $n=1/2$ when the order parameter is not
conserved and $n=1/3$ when it is locally conserved.

Dynamical scaling and dynamical universality can be combined to give a
single, time-independent description of the statistical structure of
phase-separating materials. Since $R(\tau) = [B \tau ]^n$, a
dimensionless scaled wave vector $q = k R(\tau) = k [B \tau]^n$ can be
defined and the structure factor can be collapsed onto a master curve
\begin{equation}
F'_1\left(q\right) =
[B \tau]^{-nd} S\left(k,\tau\right) \;,
\end{equation}
where $d$ is the spatial dimension of the system. Equivalently, a
scaled time
$t = q^{1/n} = [k R(\tau)]^{1/n} = k^{1/n} B \tau $ gives the master curve
\begin{equation}
\label{eq:sscale}
F_1\left(t\right)=k^{d}S\left(k,\tau\right) \;,
\end{equation}
where $F_1(t)=q^d F'_1(t^n)$. Incoherent scattering experiments are
frequently used to measure the average scattering associated with the
structure factor.

\subsection{Scaling of two-time functions}
\label{sec:two-time}

For individual speckles, time-dependent fluctuations around the
average intensity are characterized by the two-time intensity
covariance function,
\begin{eqnarray}
\label{eq:cov}
{\rm Cov_k}({\bf k},\tau_1,\tau_2) 
&=& \Big\langle I({\bf k},\tau_1)I({\bf k},\tau_2)\Big\rangle 
    - \Big\langle I({\bf k},\tau_1)\Big\rangle\Big\langle I({\bf k},\tau_2) 
      \Big\rangle \nonumber\\
&=& \Big\langle \hat\psi({\bf k},\tau_1)\hat\psi^*({\bf k},\tau_1) 
                \hat\psi({\bf k},\tau_2)\hat\psi^*({\bf k},\tau_2) 
    \Big\rangle \nonumber\\
& & - \Big\langle \hat\psi({\bf k},\tau_1)\hat\psi^*({\bf k},\tau_1) 
    \Big\rangle 
    \Big\langle \hat\psi({\bf k},\tau_2)\hat\psi^*({\bf k},\tau_2) 
    \Big\rangle 
\;.
\end{eqnarray}
Following the treatment in Ref.~\cite{BROW97A}, we anticipate that the
joint probability density of the complex variables $\hat\psi({\bf
k},\tau_1)$ and $\hat\psi({\bf k},\tau_2)$ is Gaussian.  The Gaussian
nature of the joint probability density, and thus of the marginal
probability density for each $\hat{\psi}({\bf k},\tau)$, is
essentially a consequence of the central-limit theorem. It may or may
not hold, regardless of the probability density of the order parameter
in real space $\psi({\bf r},\tau)$. Specifically, it does not depend
on whether $\psi({\bf r},\tau)$ can be derived from an underlying
Gaussian auxiliary field, an assumption which is customarily referred
to as ``the Gaussian approximation.'' In this paper we are only
concerned with the former approximation.

It is possible to construct a set a variables whose joint probability
density is not Gaussian even though the marginal probability densities
are \cite{KENDALL}, so the additional constraint that the joint
probability density be Gaussian is an important one. The essential
aspect of this assumption is that the fourth moment which occurs in
Eq.~(\ref{eq:cov}) reduces to products of second moments according to
Wick's theorem. For this reason we refer to the approximation as the
``Gaussian decoupling approximation.'' For ${\bf k} \neq 0$ the
covariance becomes equal to the square of the two-time structure
factor \cite{BROW97A},
\begin{equation}
\label{eq:equality}
{\rm Cov_k}({\bf k},\tau_1,\tau_2)=S^2({\bf k},\tau_1,\tau_2) 
\;.
\end{equation}
The two-time structure factor is defined as 
\begin{equation}
\label{eq:sk12meas}
S({\bf k},\tau_1,\tau_2) 
\equiv \Big\langle \hat\psi({\bf k},\tau_1)\hat\psi^*({\bf k},\tau_2) 
\Big\rangle \;,
\end{equation}
but it is also proportional to the Fourier transform of the two-point,
two-time order-parameter correlation function
\begin{equation}
\label{eq:skfromcr}
S({\bf k},\tau_1,\tau_2)  = 
\int d {\bf r} e^{i {\bf k \cdot r}} C({\bf r}, \tau_1, \tau_2 ) 
\;,
\end{equation}
where the correlation function is defined by
\begin{equation}
C({\bf r},\tau_1,\tau_2) =
\langle \psi({\bf 0},\tau_1) \psi({\bf r},\tau_2) \rangle
\;.
\end{equation}

Similar to the case for the one-time structure factor, dynamical
scaling also applies to the two-time structure factor. For isotropic
media, the latter can be related to a master function of the 
two scaled times,
\begin{equation}
\label{eq:sk12scale}
F_2(t_1,t_2)=k^{d}S(k,\tau_1,\tau_2) \;.
\end{equation}
Since it involves the Fourier transform of the material at two
different times, the two-time structure factor cannot be measured
directly in a scattering experiment.  However, its relationship to the
two-time intensity covariance indicates that it can be inferred from
the fluctuations of speckles around their average intensities.  In
analogy with the scaling of the two-time structure factor, a scaled
speckle covariance
\begin{equation}
\label{eq:covscaling}
{\rm Cov}(t_1,t_2)=k^{2d}{\rm Cov_k}(k,\tau_1,\tau_2)
\end{equation}
can be defined. Then
\begin{equation}
\label{eq:covscaleF}
{\rm Cov}(t_1,t_2) = F_2^2(t_1,t_2)
\;,
\end{equation}
provided the relationship between the speckle covariance and the
two-time structure factor, Eq.~(\ref{eq:equality}), is valid.

The asymptotic properties of the two-time structure factor can be
found using quite general arguments that should apply for many
universality classes. They are most concretely expressed by focusing
on the geometry of the moving interfaces, a schematic view of which is
provided in Fig.~\ref{fig:PhiAB}. The top and bottom of the figure
remain inside the domains of the two different phases, respectively,
while the shaded region changes phase between $\tau_1$ and
$\tau_2$. The edges of the shaded region represent the interface
between the domains at $\tau_1$ and $\tau_2$; the interface at
$\tau_2$ is flatter. Roughly halfway between these interfaces, the
dashed line represents the ``mean interface,'' which defines a mean
domain size, $[B \bar{\tau}\,]^n,$ where
$\bar{\tau}=(\tau_2+\tau_1)/2.$ Lengths are scaled by this mean domain
size. Using this specific choice of scaling length, along with the
ansatz that the times appear as the ratio $\tau_2/\tau_1$, the
relationship between the two-time order-parameter correlation function
and its master curve can be written
\begin{equation}
C\left({\bf r},\tau_1,\tau_2\right) = 
H\left({\bf r}/[B\bar{\tau}]^n,\delta
\tau/\bar{\tau}\,\right)
\;,
\end{equation}
where $\delta \tau = | \tau_2 - \tau_1 |.$ 

The master curve for the two-time structure factor can be found using
Eq.~(\ref{eq:skfromcr}).  When the correlation function is isotropic,
this can be reduced to a radial integral that depends on the
scaled-time variables $\delta t = | t_2 - t_1 | $ and $\bar{t} = (t_1
+ t_2)/2$;
\begin{equation}
\label{eq:smallFT}
F_2 \left(\delta t, \bar{t}\, \right) = (2\pi)^{d/2} \bar{t}\,^{n(d/2+1)}
\int_0^{\infty} du \, u^{d/2} J_{d/2-1} \left( u \, \bar{t}\,^n \right)
H \left( u , \delta t / \bar{t}\, \right)
\;,
\end{equation}
where $J_\nu$ is a Bessel function of the first kind of order $\nu.$

The small-$\bar{t}$
and large-$\bar{t}$ behaviors of $F_2$ can both be found using elements of the
picture described here.

\subsubsection{Two-time scaling for small $\bar{t}$}
\label{sec:smallt}

The small-$\bar{t}$ behavior of Eq.~(\ref{eq:smallFT}) is isolated by
Taylor expanding the Bessel function,
\begin{equation}
\label{eq:smalltexpand}
F_2 \left(\delta t, \bar{t}\, \right) = 2 \pi^{d/2} \bar{t}\,^{n d}
\sum_{j=0}^{\infty} (-1)^j \bar{t}\,^{2 n j} 
h_{2j} \left(\delta t/\bar{t}\,\right)
\;,
\end{equation}
where $h_{2j}(\delta t/\bar{t}\,)$ is defined through the integration
over the scaled distance $u$,
\begin{equation}
h_{2j} \left(\delta t/\bar{t}\,\right) =
\frac{\int_0^{\infty} du \, u^{2j+d-1}
       H \left(u,\delta t/\bar{t}\,\right)}
     {4^j\,j!\, \Gamma\left(j+d/2\right)}
\;.
\end{equation}
At small wave vector the structure factor can be shown to obey a power
law $S \propto k^{\beta(n)}$, or equivalently $F_2 \propto
\bar{t}\,^{n(\beta(n)+d)}$. The value is $\beta ( 1/2 ) = 0$ for
model~A \cite{OHTA82} and believed to be $\beta ( 1/3 ) = 4$ for
model~B \cite{YEUN88}. For model B, this indicates that
$h_0(0)=h_2(0)=0,$ and, consequently, that the leading term is the one
containing $h_4(\delta t/\bar{t})$. The Taylor expansion is well
approximated by its leading term when that term is much larger than
the one containing $h_6$. This condition is equivalent to
\begin{equation}
\label{eq:smalltvalid}
\bar{t}\,^{2n} \ll
12 \left( \frac{d}{2}+2 \right)
\frac{\int_0^{\infty} du \, u^{d+3} H(u,\delta t/\bar{t}\,)}
     {\int_0^{\infty} du \, u^{d+5} H(u,\delta t/\bar{t}\,)}
\;,
\end{equation}
where the ratio of the integrals is a function of $\delta t/\bar{t}.$
For $\bar{t}$ small enough, then, the scaling of the two-time
structure factor depends on $\delta t$ only through $\delta
t/\bar{t}$. The analogous result for model~A (which follows with
$h_0(0)>0$) was already obtained explicitly in Ref.~\cite{BROW97A} for
the Yeung-Jasnow (YJ) \cite{YEUN90} correlation function. This
small-$\bar{t}$ behavior agrees with our numerical results for both
model~A \cite{BROW97A} and model~B (see Sec.~\ref{sec:results}).

\subsubsection{Two-time scaling for large $\bar{t}$}
\label{sec:bigt}

When Eq.~(\ref{eq:smalltvalid}) is not satisfied, a large number of
terms is required in Eq.~(\ref{eq:smalltexpand}) to obtain an accurate
estimate of $F_2(\delta t, \bar{t}\,).$ Since these terms contain
products of powers of $\bar{t}$ and $\delta t/\bar{t},$ $F_2(\delta t,
\bar{t}\,)$ should not be expected to depend on $\delta t$ only
through $\delta t/\bar{t}$ for large $\bar{t}.$ In Ref.~\cite{BROW97A}
this was demonstrated explicitly for model~A using the YJ
\cite{YEUN90} analytic approximation for the correlation function. The
result presented here is based on simple geometric arguments, which
should be valid for all universality classes with a scalar order
parameter, including both model~A and model~B.

At large wave vectors the one-time structure factor for a two-phase
system with sharp, randomly oriented interfaces is well described by
Porod's law \cite{POROD}, $S \propto k^{-(d+1)}$.  Here the
proportionality constant includes the interface area per unit volume
(the specific surface) at time $\tau$, $A(\tau) \propto 1/R(\tau)$.
In this limit the scattering probes the correlation function at
lengths $r \ll R(\tau)$, where the normalized equal-time correlation
function decreases linearly with scaled distance $A(\tau) r \propto
r/R(\tau)$ \cite{DEBY57,TOMI84,TOMI90},
\begin{equation}
\label{eq:1timeC}
\frac{C({\bf r}, \tau , \tau)} 
{\langle \psi^2(\tau) \rangle }
= 
1 - C_1  \frac{r}{[B \tau]^n} 
\;.
\end{equation}
Here $C_1$ is a nonuniversal constant.  The mean-square order
parameter $\langle \psi^2 (\tau) \rangle$ is independent of $\tau$ if
the two equilibrium values of $\psi$ are equal in magnitude or the
volume fractions of the two phases are constant, as they are in Model
B, and we will simply denote it $\langle \psi^2 \rangle$.

\begin{figure}[tb]
\vskip 6in
\includegraphics{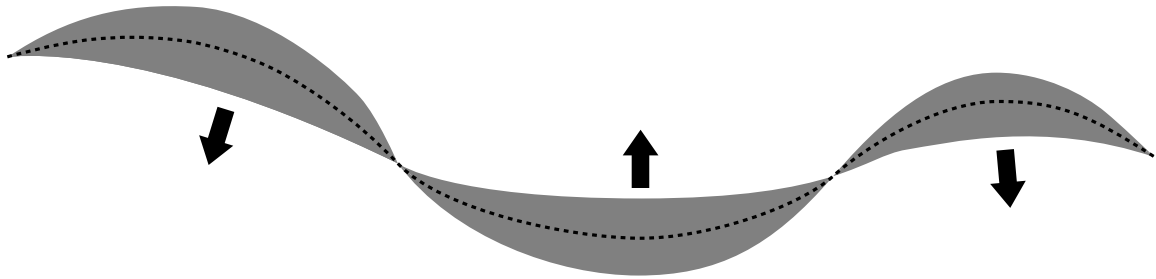}
\bigskip
\bigskip
\caption[]{Schematic illustration of the moving-interface model. The
edges of the shaded region represent the interface at times $\tau_1$
and $\tau_2$, and the arrows indicate the directions of motion. The
dashed curve is the mean interface. The shaded region represents the
volume fraction $\phi_{\Delta}$, which changes phase between $\tau_1$
and $\tau_2$.}
\label{fig:PhiAB}
\end{figure}

The behavior for small, but nonzero, time differences can be deduced
from the moving-interface model. The normalized two-time
autocorrelation function $C({\bf 0}, \tau_1 , \tau_2 ) / \langle
\psi^2 \rangle $ is reduced from unity by an amount proportional to
$\phi_{\Delta}$, the fraction of the system volume which is occupied
by different phases at $\tau_1$ and $\tau_2$, {\em i.e.} the shaded
volume in Fig.~\ref{fig:PhiAB}.  In the sharp-interface picture used
here, this is an exact result, obtained by simple probability
arguments. The proportionality constant is easily calculated but not
very useful.

For small time differences in a moving-interface model, one can
construct a coordinate system based on the mean interfacial position,
represented by the dashed curve in Fig.~\ref{fig:PhiAB}, and a
coordinate locally perpendicular to it, $\Delta$.  The specific
surface of the mean interface is approximately $\bar{A} \approx
[A(\tau_1) + A(\tau_2)]/2$, and the average distance between the other
interfaces is $\langle | \Delta | \rangle$. In this approximation
$\phi_{\Delta} \approx \bar{A} $ $\langle | \Delta | \rangle$.  Using
the scaling assumption that $R(\tau) = [ B \tau]^n \propto 1/A(\tau)$
is the only length scale characterizing the system at time $\tau$, we
obtain
\begin{equation}
\label{eq:timescaling}
\langle | \Delta | \rangle 
\propto 
|R(\tau_2) - R(\tau_1)| 
\approx 
\left.\frac{d R(\tau)}{d \tau} \right|_{\tau = \bar{\tau}} \delta \tau 
=  
n B^n \frac{\delta \tau}{ \bar{\tau}^{1-n}} 
\equiv
n B^n \zeta 
\;. 
\end{equation}
The dimensionless proportionality constant between $\langle | \Delta |
\rangle$ and $|R(\tau_2) - R(\tau_1)|$ is expected to depend on the
particular dynamic model and the dimensionality $d$. The relation
$|R(\tau_2) - R(\tau_1)| \approx n B^n \zeta$ is a Taylor expansion
valid for small $\zeta$. These results can be combined to give
\begin{equation}
\frac{C({\bf 0},\tau_1,\tau_2) }
{\langle \psi^2 \rangle }
= 
1 - \frac{n C_1 C_2 \zeta}{\bar{\tau}\,^n} 
\;,
\end{equation}
where $C_2$ contains the proportionality constants from
Eq.~(\ref{eq:timescaling}).

On length scales $\langle | \Delta | \rangle \ll r \ll 1/\bar{A}$ the
interface should appear sharp, so the asymptotic correlation function
should be described using the mean domain size, {\em i.e.} by
Eq.~(\ref{eq:1timeC}) with $\tau^{-n}$ replaced by
$\bar{\tau}\,^{-n}$.  The crossover between the small-$r$ and
intermediate-$r$ limits can be described in terms of a scaling
function $G(x)$ as
\begin{mathletters}
\begin{equation}
\label{eq:scalCr12} 
\frac{C({\bf r} , \delta \tau , \bar{\tau}) }
{\langle \psi^2 \rangle }
= 
1 - \frac{n C_1 C_2 \zeta}{\bar{\tau}^n} \ 
G\left( \frac{r}{n C_2 B^n \zeta} \right) 
\;, 
\end{equation}
with the asymptotic behavior
\begin{equation}
\label{eq:scalfG} 
G(x) \sim 
\left\{ 
\begin{array}{ll} 
1 & \mbox{for $x \ll 1$} \\
x & \mbox{for $x \gg 1$} 
\end{array}
\right.
\;.
\end{equation}
\label{eq:scalfeq}
\end{mathletters}
Thus, $C({\bf r}, \delta \tau, \bar{\tau})$ depends on $r$ only
through the dimensionless scaling combination $r/[n C_2 B^n \zeta]$.
As seen from the derivation, this result is general and applies to any
moving-interface model.

The two-time structure factor in the large-$k$ limit is obtained from
Eq.~(\ref{eq:scalfeq}) using the same formula for the $d$-dimensional
Fourier transform of an isotropic function as in
Eq.~(\ref{eq:smallFT}),
\begin{equation}
\label{eq:spherFT}
S(k,\tau_1,\tau_2) 
 = - k^{-d} 
\langle \psi^2 \rangle 
(2\pi)^{d/2} 
\frac{n C_1 C_2 \zeta}{\bar{\tau}^n} 
\int_0^\infty d u \, u^{d/2}\,J_{d/2 - 1}(u)\, 
G\left( \frac{u}{n C_2 k B^n \zeta} \right)
\;,
\end{equation}
where we have removed from $C({\bf r}, \delta \tau , \bar{\tau})$ the
constant term which only contributes to a $\delta$ function at $k =
0$.  Defining the new dimensionless scaling variable $z = k B^n \zeta
\equiv \delta t / \bar{t}\,^{1-n}$, we obtain the asymptotic scaling
form for the two-time structure factor in the large-$\bar{t}$ limit:
\begin{equation}
\label{eq:tnF2}
\Phi_d(z) \equiv \frac{ \bar{t}\,^n F_2(z,\bar{t}) } 
{ \langle \psi^2 \rangle C_1 }
 = - (2\pi)^{d/2} n C_2 z 
\int_0^\infty d u \, u^{d/2}\,J_{{d/2}-1}(u)\, 
G\left( \frac{u}{n C_2 z} \right)
\;.
\end{equation}
This asymptotic scaling function depends on $k$, $\tau_1$, and
$\tau_2$ only through $z$.  If the correlation function vanishes
sufficiently smoothly for large $r$, standard expansions of Fourier
transforms \cite{TOMI90,ERDE56} indicate that the asymptotic large-$k$
behavior of the structure factor is determined completely by the
real-space behavior of the correlation function for small $r$.  For $z
= 0$ it reduces to the single-time scaling function \cite{TOMI90}
\begin{equation}
\label{eq:tnF2z0}
\Phi_d(0) = \frac{ \bar{t}\,^n F_2(0,\bar{t}) } { C_1 }
 =  2^d \pi^{(d-1)/2} \Gamma \left( \frac{d+1}{2} \right) 
\;,
\end{equation} 
which is consistent with the amplitude of the large-$k$ Porod tail of
the structure factor ($\Phi_2(0)=2\pi$ and $\Phi_3(0)=8\pi$). As $z$
increases, the integral in Eq.~(\ref{eq:tnF2}) tends towards the
Fourier transform of a constant, which vanishes for all nonzero $k$.
Thus, $\Phi_d(z)$ decreases from its maximum value at $z = 0$ towards
zero as $z$ increases.

In Ref.~\cite{BROW97A} the scaling behavior expressed in
Eq.~(\ref{eq:tnF2}) was explicitly obtained for model~A, but a
nonrigorous scaling argument suggested it should hold for other
dynamical models as well. This was recently confirmed by XIFS
experiments in a sodium borosilicate glass undergoing spinodal
decomposition \cite{MALI98}.  This is a two-phase system with a
locally conserved scalar order parameter appropriately described by
model~B.

To predict the form of $\Phi_d(z)$ in further detail it is necessary
to know how $G(x)$ interpolates between the limiting behaviors given
by Eq.~(\ref{eq:scalfG}). A detailed calculation based on an extension
of an approach used by Tomita to calculate equal-time correlation
functions \cite{TOMI84,TOMI90}, indicates that $G(x) - x \propto
x^{-1}$ as $x \rightarrow \infty$ and $G(x) - 1 \propto x^{2}$ as $x
\rightarrow 0$ for moving-interface models \cite{INPREP}.  A
convenient analytic form with this behavior is $G(x) = \sqrt{1 +
x^2}$. Choosing $v =n C_2 z$ and the substitution $u^2 + v^2 =
(vy)^2$, the integral $I(v) \equiv \Phi_d(v/n C_2)$ is converted to
\cite{KORNISS}
\begin{eqnarray}
I(v) 
&=& 
- (2 \pi)^{d/2} \int_0^\infty du \ u^{d/2} 
  J_{d/2-1}(u) \ \sqrt{u^2 + v^2} 
\nonumber\\
&=& 
- (2 \pi)^{d/2} v^{d/2+2} \int_1^\infty dy \ y^2 
  \left( \sqrt{y^2-1} \right)^{d/2-1} 
  J_{d/2-1} \left(v \sqrt{y^2-1} \right) 
\;.
\label{eq:A21}
\end{eqnarray}
Asymptotic convergence is guaranteed by the discussion following
Eq.~(\ref{eq:tnF2z0}), and it is possible to use the approach of
evaluating a generating function for the desired integral as an Abel
limit \cite{KORNISS,WONG89}
\begin{equation}
I(v) 
= 
- (2 \pi)^{d/2} v^{d/2+2} 
\lim_{\alpha \rightarrow 0} \frac{\partial^2}{\partial \alpha^2} 
\int_1^\infty dy \ e^{-\alpha y}\left( \sqrt{y^2-1} \right)^{d/2-1} 
  J_{d/2-1} \left( v \sqrt{y^2-1} \right) 
\;.
\label{eq:A22}
\end{equation}
This integral can be found in tables \cite{GandR}, and 
differentiation followed by taking the limit leads to 
\begin{equation}
\label{eq:Phizsqrt}
\Phi_d(z) 
 = \Phi_d(0) 
\frac{\left(n C_2 z \right)^{(d+1)/2} K_{(d+1)/2}\left(n C_2 z \right)} 
{2^{(d-1)/2} \Gamma \left( (d+1)/2 \right)}
\;,
\end{equation}
where $K_\nu$ is a modified Bessel function of the second kind. For
$d=3$ this can be numerically evaluated without difficulty, while for
$d = 2$ the expression simplifies further to \cite{AandS}
\begin{equation}
\label{eq:Phizsqrt2}
\Phi_2(z) 
 = 2 \pi \left( 1 + n C_2 z \right) e^{-n C_2 z} \;.
\end{equation}
The same result was obtained in Ref.~\cite{BROW97A}, using the full YJ
\cite{YEUN90} correlation function for model~A (see Appendix). The
advantage of the present approach is that it demonstrates that
$\Phi_d(z)$ is manifestly independent of the large-$r$ behavior of the
correlation function, requiring only that it converges sufficiently
smoothly to ensure the equality of the Abel limit and the integral.

\section{Numerical Procedure}
\label{sec:numerics}

In phase-separating systems no material is created or destroyed. The
order parameter is locally conserved, and evolution occurs by
diffusion along chemical-potential gradients.  The Cahn-Hilliard-Cook
model \cite{CAHN58,COOK70} is a convenient description for the
dynamics of a conserved scalar order parameter.  The thermodynamics of
the system are described by the Ginzburg-Landau-Wilson free energy
\cite{GUNT83},
\begin{equation}
\label{eq:GLWFE}
{\cal F}[\psi({\bf r},\tau)]=\int d{\bf r} 
\Bigg[-\frac{a}{2}\psi^2({\bf r},\tau )
+\frac{u}{4}\psi^4({\bf r},\tau)
+\frac{c}{2}|{\bf\nabla}\psi({\bf r},\tau)|^2 \Bigg] \;.
\end{equation}
For $a>0$, the first two terms of the integrand create a bistable
local potential energy. The last term represents the surface tension
between domains in which $\psi$ has opposite sign.  The dynamics are
implemented using the Langevin equation
\begin{equation}
\frac{\partial \psi({\bf r},\tau)}{\partial \tau} = 
M \nabla^2 \frac{\delta {\cal F}[\psi({\bf r},\tau)]}{\delta \psi({\bf r},\tau)}
+ \eta({\bf r},\tau) \;.
\label{eq:CHC}
\end{equation}
The first term on the right-hand side represents deterministic
relaxation of the chemical potential; the Laplacian expresses the
local conservation constraint.  Processes operating at shorter time
and length scales are considered thermal noise and are modeled by the
random variable $\eta$, whose intensity is given by a
fluctuation-dissipation theorem \cite{GUNT83}.  We neglect $\eta$
because the most important sources of noise here are the initial
conditions, which give the random domain morphology at early times.

For the symmetric mixtures considered here, the parameters can be
eliminated by appropriately normalizing the time, length, and
concentration scales\cite{BROW97A,GRAN85,SCALED-VAR}.  Using the same
names for the new variables, the resulting dynamical equation is
\begin{equation}
\frac{\partial\psi({\bf r},\tau)}{\partial \tau}
  = -\frac{1}{2} \nabla^2 \left[ \left( 1 + \nabla^2 \right) \psi({\bf r},\tau) 
  - \psi^3({\bf r},\tau) \right] \;.
\label{eq:EOM} 
\end{equation}
In particular, this yields $\psi = \pm 1$ for the equilibrium values of the 
order parameter. 

We have simulated this model on a square lattice, with $\Delta r$=1
and $L_x=L_y=L=512$, using a simple Euler integration scheme with
$\Delta \tau=0.05$.  The initial condition was implemented by choosing
random values uniformly distributed between $\pm 0.1$ for each point
on the lattice.  Measurements were made every $50$ time units out to a
maximum of $\tau=4000$.  As can be seen from Fig.~\ref{fig:config},
the domains are much smaller than the system size, even at this latest
time.  Finite-size effects are therefore not expected to affect the
order-parameter dynamics.  Indeed, no deviations from the expected
behavior of the characteristic length are observed in the simulations.

The other numerical procedures are identical to those described in
Ref.~\cite{BROW97A}. The Laplacian was implemented using
eight-neighbor discretization \cite{OONO87b,TOMI91}, and the magnitude
of the wave vector $k({\bf k})$ was defined in a manner consistent
with that Laplacian. In addition, the Fourier transform of the
hardened order parameter, ${\rm sgn}[\psi({\bf r},\tau)]$, was used to
minimize the effect of finite interface width on the scattering
intensity.

\section{Results}
\label{sec:results}

The Gaussian decoupling approximation that leads to
Eq.~(\ref{eq:equality}) has not been directly justified for
phase-ordering systems. Gaussian variables are often assumed in the
context of the central limit theorem, where it can be argued that many
independent random variables with finite variance contribute
additively. Since $R(\tau)$ represents the average domain size, at any
given time there are on the order of $(L/R(\tau))^d$ independent
domains in a system of edge-length $L$. This number can be quite
large, but since the domains interact as they grow, with material
diffusing along chemical-potential gradients, the applicability of the
central limit theorem is not obvious. In our previous study of a
system with nonconserved order parameter \cite{BROW97A}, the
decoupling was justified {\em a posteriori} from the numerical
results. Similar justification is presented here for numerical
integration of the equation of motion for the conserved order
parameter. When discussing the numerical results, we have taken $B=1$
for convenience.

\begin{figure}[tb]
\vskip 2.85in
\includegraphics{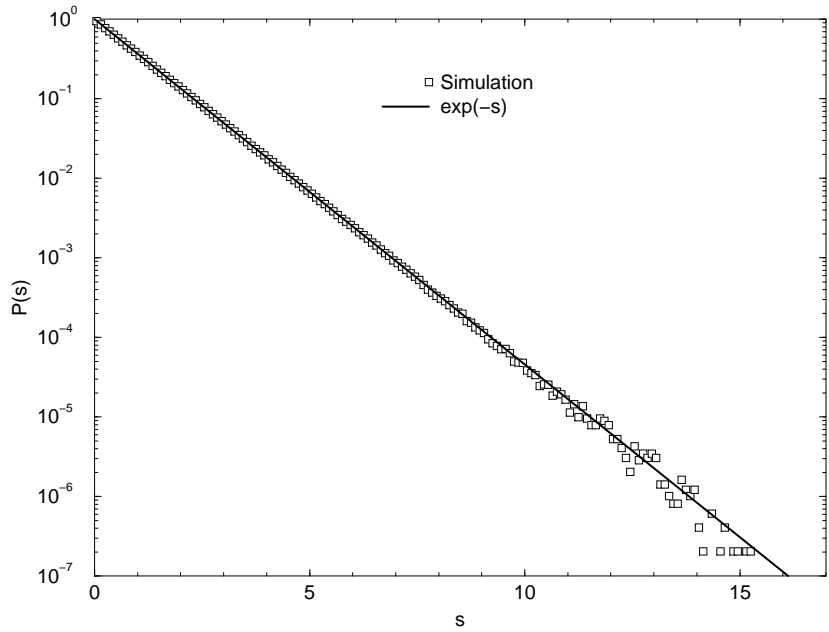}
\bigskip
\smallskip
\caption[]{Probability density of the normalized scattering intensity
$s(k,\tau)=I(k,\tau)/S(k,\tau)$, binned for all times and $0.08<k({\bf
k})<0.75$. The histogram is exponential for all $s$ observed here,
indicating the Gaussian decoupling approximation used to equate the
squared two-time structure factor and the speckle covariance is a
reasonable approximation at the one-time level.}
\label{fig:expint}
\end{figure}

One way to test the validity of the assumption at the single-time
level is looking at the probability density of the speckle
intensities. Namely, Eq.~(\ref{eq:equality}) with $\tau_1 = \tau_2$ is
satisfied if the normalized scattering intensity, $s({\bf
k},\tau)=I({\bf k},\tau)/S({\bf k},\tau)$, has an exponential
probability density,
\begin{equation}
P(s)=\exp{(-s)} \;,
\end{equation}
that is independent of $({\bf k},\tau)$. A single histogram was
constructed for the normalized intensity for all times and for $0.08 <
k({\bf k}) < 0.75$ \cite{k-RANGE}.  The intensity at a particular
$(k,\tau)$ was normalized by $S(k,\tau)$ estimated by circular
averaging only for the {\em same} quench experiment. The lower $k$
cutoff was chosen so that at least $20$ speckles contributed to that
circular average, while the upper cutoff was set using results for
$\tau_1 \neq \tau_2$ (discussed below).  Nearly $10^8$ samples
contribute to the histogram in Fig.~\ref{fig:expint}, and the
agreement with the exponential form is remarkable. Other histograms
were constructed to check for $k$ and $\tau$ dependence, but are not
shown here. Wave vector dependence was investigated by sampling from
narrow rings of constant $k$ at all times, and time dependence was
checked by sampling all $k$ between the cutoffs at single times. The
probability density for the normalized intensity does not appear to
depend on these variables. A similar check of the Gaussian decoupling
assumption was made by Shinozaki and Oono \cite{SHIN93} for a
cell-dynamical simulation with conserved order parameter.

\begin{figure}[tb]
\vskip 2.85in
\includegraphics{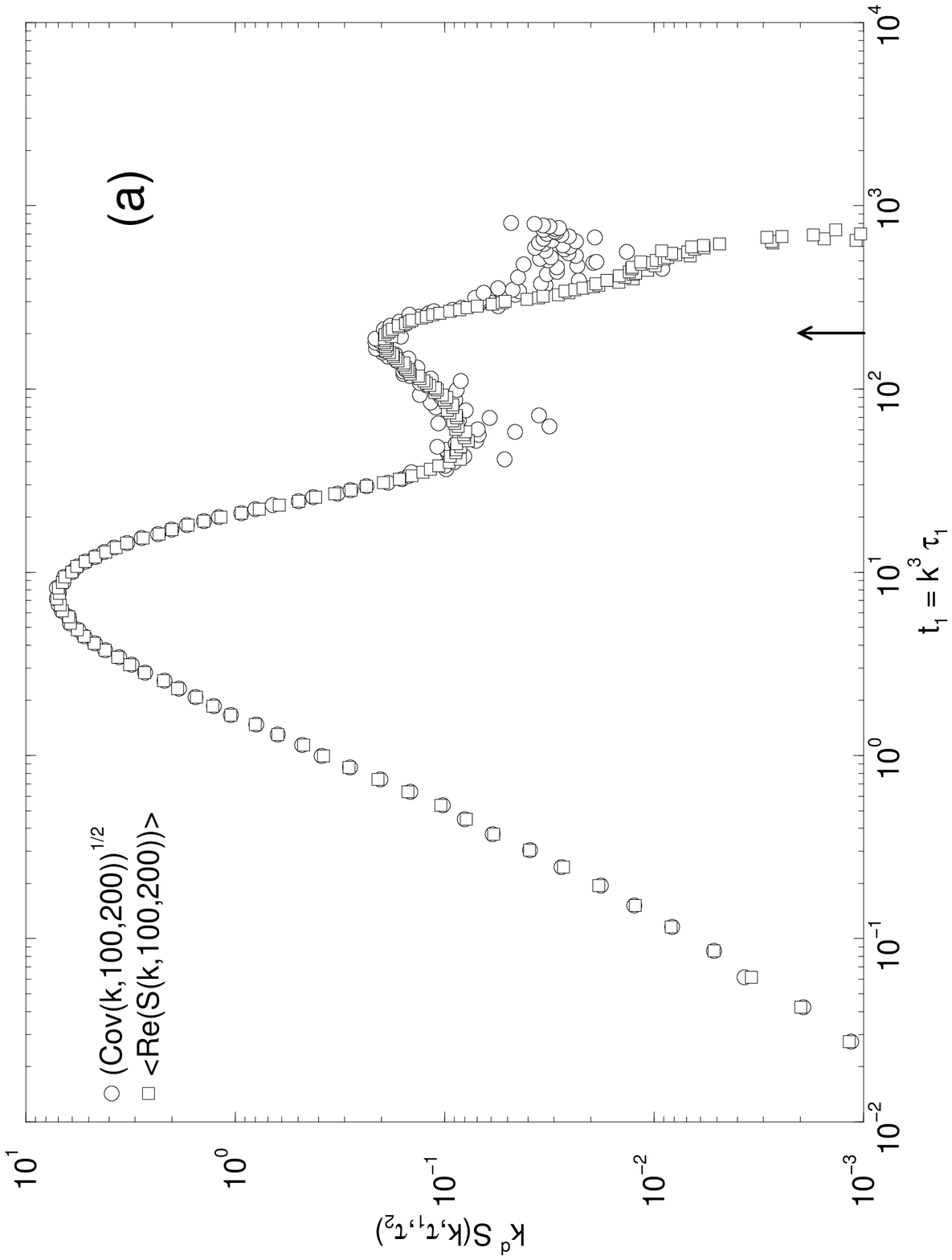}
\vskip 0.5in
\vskip 2.85in
\includegraphics{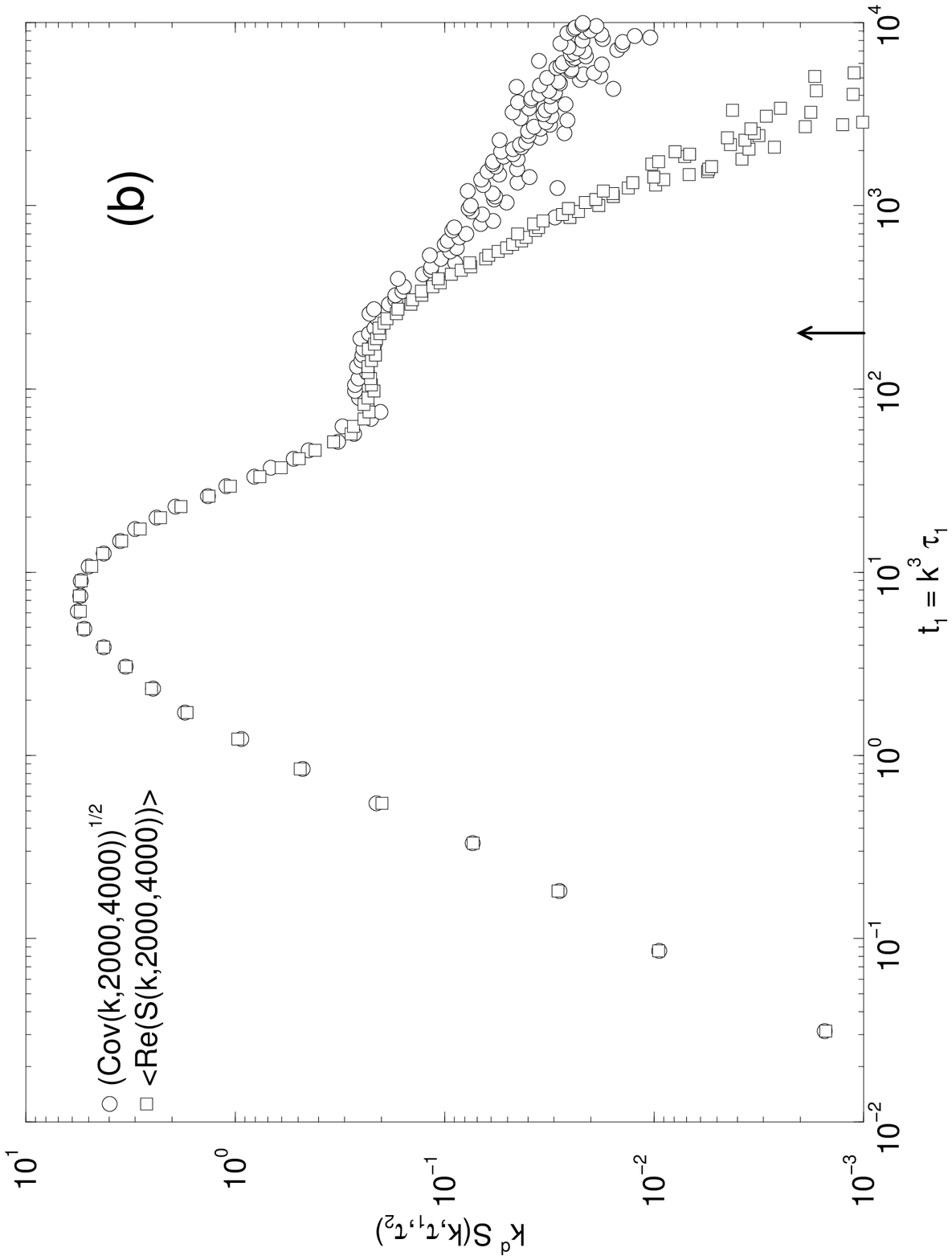}
\bigskip
\caption[]{Comparison of the two-time structure factor
$S(k,\tau_1,\tau_2)$ (squares) and the square-root of the covariance,
$\sqrt{{\rm Cov_k}(k,\tau_1,\tau_2)}$ (circles) for (a) $\tau_1=100$,
$\tau_2=200$ and (b) $\tau_1=2000$, $\tau_2=4000$. These examples
serve as a direct test of Eq.~(\ref{eq:equality}). The agreement is
good for small wave vectors, suggesting that two-time correlation
functions for real systems can be measured via speckle
experiments. The arrows indicate the scaling variable $z \approx 4.5$
where the Gaussian decoupling for two-time correlations breaks down in
these simulations.  For $\tau_2/\tau_1$=2, as in this figure, this
corresponds to $k^3\tau_1 \approx 190.$}
\label{fig:skt1t2}
\end{figure}

The equality between the two-time structure factor squared and the
two-time intensity covariance implied by the Gaussian decoupling
assumption can also be tested directly. These two quantities can be
estimated from the simulations using Eq.~(\ref{eq:sk12meas}) and
Eq.~(\ref{eq:cov}), respectively, with circular averaging over ${\bf
k}$ in addition to the averaging over the quenches.  Since the
real-space correlation function is an even function of ${\bf r}$,
$S({\bf k},\tau_1,\tau_2)$ is real valued. We have verified that the
imaginary part is zero to within the accuracy of our results. The
equality is tested graphically in Fig.~\ref{fig:skt1t2} as a function
of $t_1=k^3\tau_1$ for (a) $\tau_1=100$, $\tau_2=200$ and (b)
$\tau_1=2000$, $\tau_2=4000$.  The intensity covariance and the
squared structure factor are equal within our numerical accuracy for
sufficiently small $k$. However, for $z \agt 4.5$ (marked by arrows in
both parts of Fig.~\ref{fig:skt1t2}) the covariance becomes larger
than the squared structure factor, indicating the gradual breakdown of
the Gaussian decoupling approximation for large $z$ in these
simulations. A systematic investigation of the range of validity of
the Gaussian decoupling approximation, including the possible
system-size dependence of the number of independent contributions to
the Fourier transform, is left for future study. In this paper all
quantitative scaling results are based only on data for $k({\bf k}) <
0.75$ \cite{k-RANGE}, where the squared two-time structure factor
appears to equal the speckle covariance for all times considered.

\begin{figure}[tb]
\vskip 6in
\includegraphics{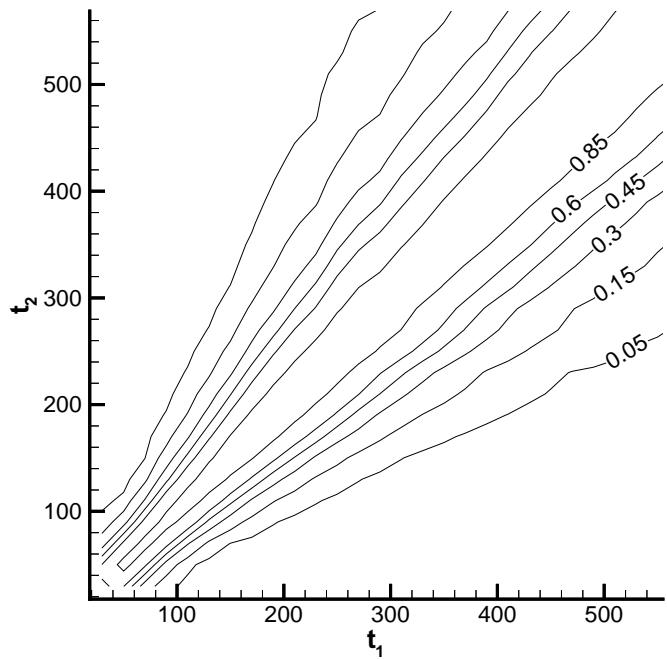}
\bigskip
\caption[]{Contour plot of the scaled speckle two-time correlation
 function ${\rm Corr}(t_1, t_2)$. The correlation decreases as one
 moves away from the line $t_1=t_2$, where it is unity by
 construction.}
\label{fig:contour}
\end{figure}

The scaling function $F_2(t_1,t_2)$, which describes both the two-time
structure factor and the related speckle covariance, depends on two
rescaled times.  A normalized analog of the scaled covariance, the
correlation function, can be defined as
\begin{equation}
\label{eq:MM}
{\rm Corr}(t_1,t_2) = 
\frac{\Big\langle I(k,\tau_1)I(k,\tau_2)\Big\rangle}
     {\Big\langle I(k,\tau_1) \Big\rangle  
      \Big\langle I(k,\tau_2) \Big\rangle } - 1
\;.
\end{equation}
Since the Gaussian decoupling approximation for scattering has been
verified for $\tau_1 = \tau_2$, this quantity is unity by construction
for $t_1=t_2$. Thus the contour plot of ${\rm Corr}(t_1,t_2)$,
Fig.~\ref{fig:contour}, is more illustrative than one for ${\rm
Cov}(t_1,t_2)$. The correlation function decays as one moves away from
the $t_1=t_2$ diagonal. Given the symmetry under exchange of $t_1$ and
$t_2$, $\bar{t}$ and $\delta t$ are more natural variables for the
scaling functions ${\rm Corr}$ and ${\rm Cov}$.  In particular,
$\bar{t}$ measures distance (in units of scaled time) along the
diagonal, while $\delta t$ measures the distance perpendicular to the
diagonal.  In Fig.~\ref{fig:contour} it is apparent that the speckle
intensity stays correlated for larger $\delta t$ as $\bar{t}$ is
increased.  As noted in Ref.~\cite{BROW97A}, all the quantities used
here are readily obtained in experiments, and the method of data
analysis should be well suited for experimental analysis as well.

\begin{figure}[tb]
\vskip 2.85in
\includegraphics{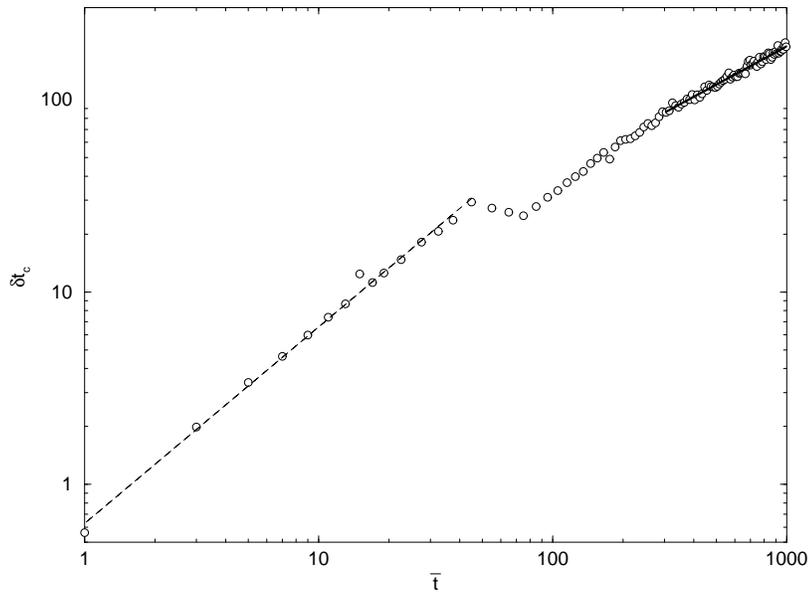}
\bigskip
\caption[]{Characteristic time difference $\delta t_c$ defined by
${\rm Cov}(\delta t_c,\bar{t}\,)=(1/2){\rm Cov}(0,\bar{t}\,)$. The
lines are least-squares fits to power-law behavior. For $\bar{t} < 50$
(dashed line) the fit gives an exponent $1.02 \pm 0.02$, which agrees
with our expectation for small $\bar{t}$.  For $\bar{t} \ge 300$
(solid line) the least-squares fit gives an exponent $0.66 \pm
0.01$. This agrees with what is expected from the large-$\bar{t}$
scaling variable $z=\delta t\,/\bar{t}\,^{1-n}$ with $1-n=2/3$, as
well as with a recent XIFS experiment [8].}
\label{fig:chartime}
\end{figure}

The asymptotic scaling predicted for large and small $\bar{t}$ can
also be tested. First, the characteristic time difference $\delta t_c$
defined by ${\rm Cov}(\delta t_c,\bar{t}\,)=\frac{1}{2}{\rm
Cov}(0,\bar{t}\,)$ can be found for fixed ranges of $\bar{t}$. The
measured $\delta t_c$, found by linear interpolation, is presented as
a function of $\bar{t}$ on a log-log scale in
Fig.~\ref{fig:chartime}. The results show obvious power law behavior
at both small and large $\bar{t}.$ A least-squares fit for $ \bar{t} <
50 $ gives an exponent of $1.02 \pm 0.02$, which agrees with our
expectation that in this regime the two-time structure factor scales
with $\delta t/\bar{t}$ (Sec.~\ref{sec:smallt}).  Least-squares
fitting for $ \bar{t} >300 $ gives an exponent of $0.66 \pm 0.01$ in
excellent agreement with the theoretical prediction of $1-n = 2/3$
expected from $z=\delta t / \bar{t}\,^{1-n}$
(Sec.~\ref{sec:bigt}). The fit also indicates that the characteristic
$\bar{t}$ for crossover between these regimes is $\bar{t}_c \sim 60$,
which corresponds to the shoulder in the structure factor (see Figs.~6
and 10).

\begin{figure}[tb]
\vskip 2.85in
\includegraphics{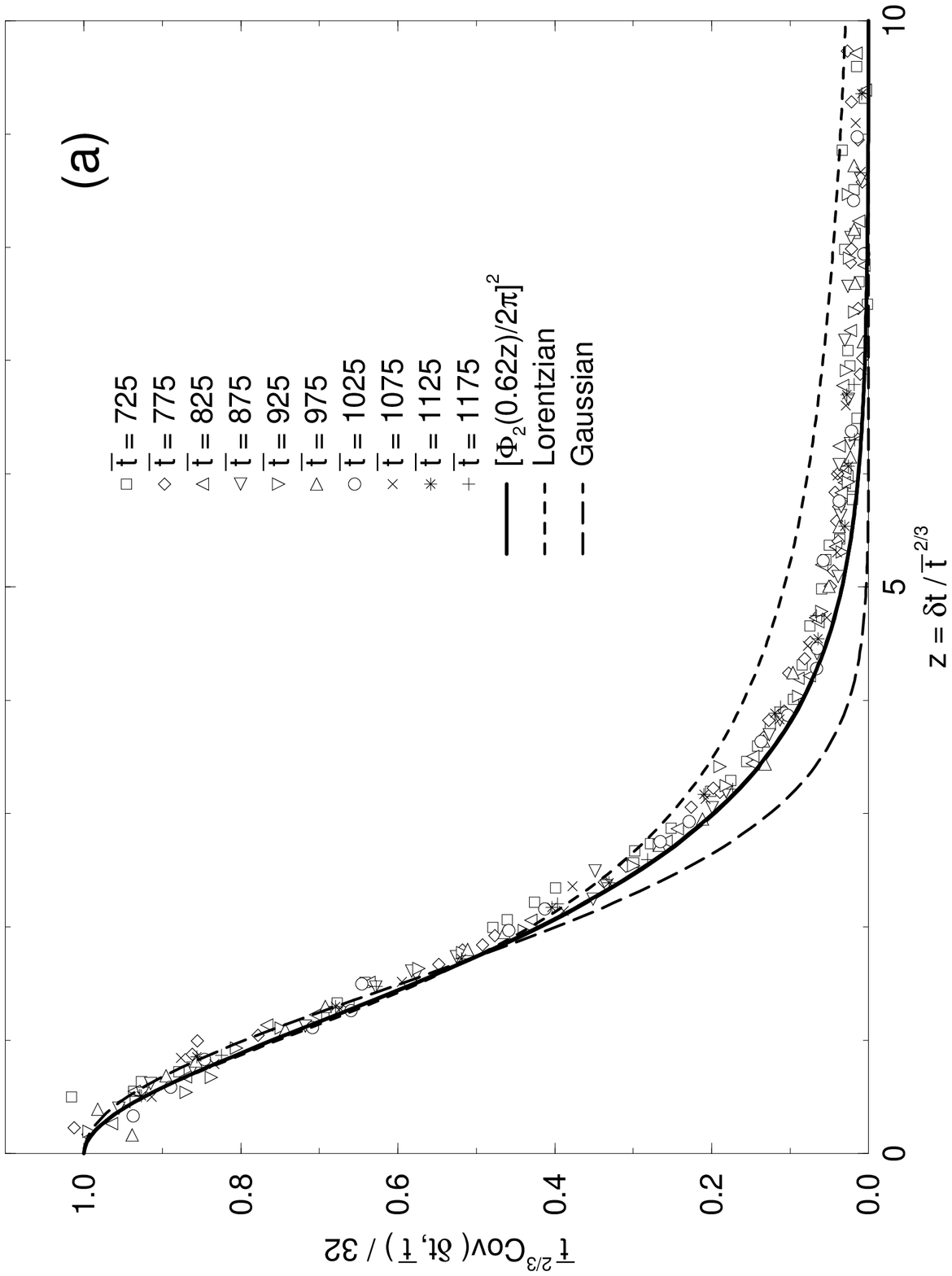}
\vskip 0.5in
\vskip 2.85in
\includegraphics{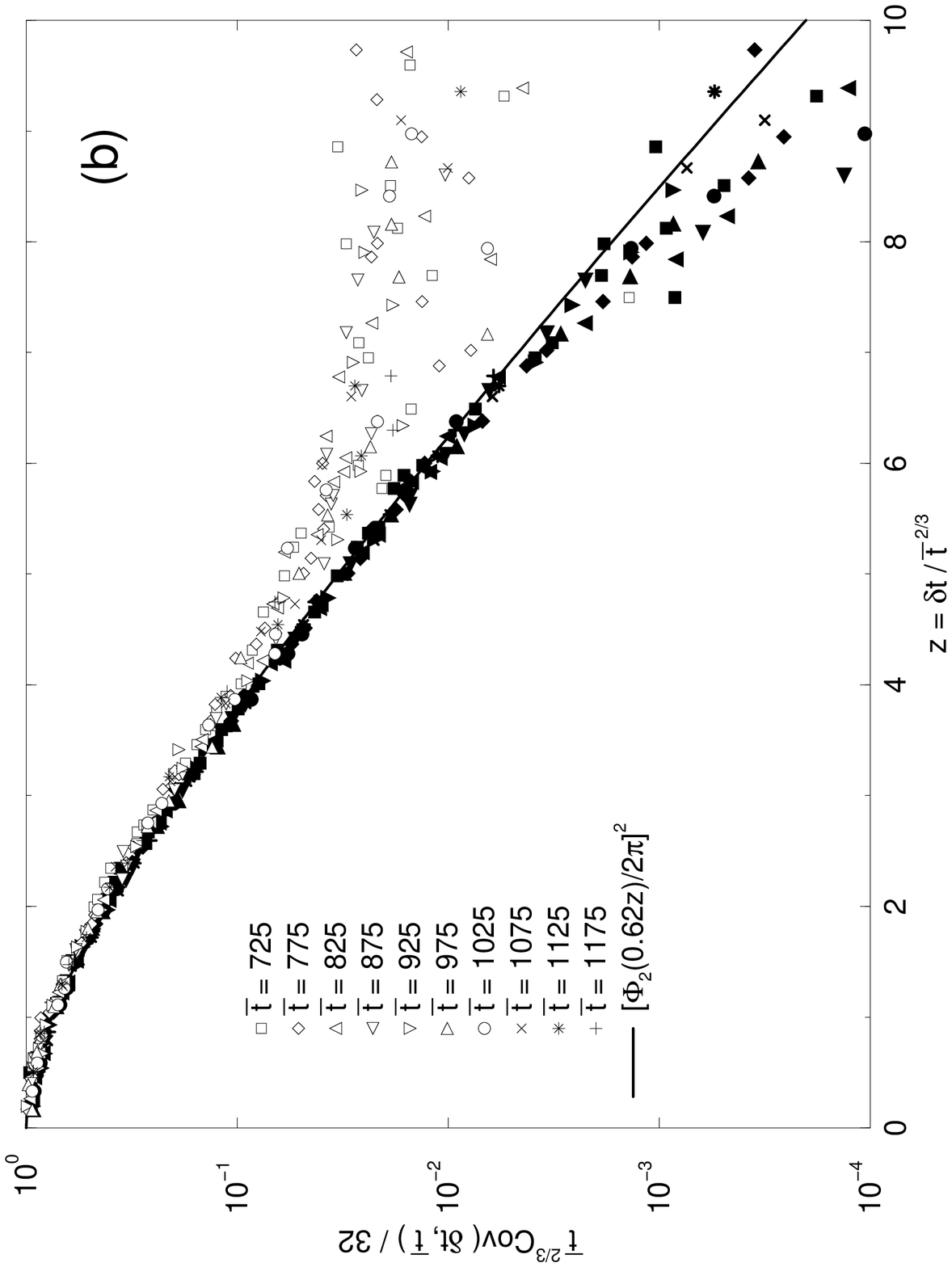}
\bigskip
\caption[]{Scaling of the covariance vs $\delta t/\bar{t}\,^{1-n}$ for
several values of $\bar{t}$ is shown.  (a) Linear scale.  (b)
Linear-log scale, where data for the two-time structure factor are also
shown (filled symbols).  The solid curve is the theoretical form,
Eq.~(27). Inspection of data indicates $(2 \pi C_1)^2 \approx 32$, and
numerical fitting to the two-time structure factor data gives $n C_2
\approx 0.62$. The data collapse is good, confirming that there is a
single master curve for the speckle covariance in the large-$\bar{t}$
limit. The deviation of the covariance data from the two-time
structure factor for $z \agt 4.5$ indicates the gradual breakdown of
the Gaussian decoupling approximation
for these simulations.}
\label{fig:covscale} 
\end{figure}

The scaling function $\Phi_d(z)$ associated with the speckle
covariance at large $\bar{t}$ is shown in Fig.~\ref{fig:covscale}(a)
for several $\bar{t} > 700$. Here the normalization was determined
from the asymptotic data where it appears that $(2 \pi C_1)^2 \approx
32$. The collapse of the data is quite good, and the agreement with
the theoretical scaling form $\Phi_2(z)$ from Eq.~(\ref{eq:Phizsqrt2})
is also excellent for $z \alt 4.5$.  These results are shown on a
linear-log scale in Fig.~\ref{fig:covscale}(b), along with the scaled
two-time structure factor data for the same $\bar{t}$ (drawn as filled
symbols).  The fitting parameter $n C_2 \approx 0.62$ was determined
from a nonlinear fit to the two-time structure factor data. Fitting to
the covariance data did not significantly increase the agreement with
$\Phi_2(z).$ This theoretical scaling form gives much better overall
agreement with the data than more conventional choices, such as a
Gaussian or Lorentzian.  Beyond $z \approx 4$ the intensity covariance
is quite small. However, as $z$ increases further it becomes larger
than the squared structure factor, which remains reasonably well
described by $\Phi_2(z)$ for all $z$ studied. This deviation signals
the breakdown of the Gaussian decoupling approximation, and is the
same deviation seen at large $t_1$ in Fig.~\ref{fig:skt1t2}.

\begin{figure}[tb]
\vskip 2.85in
\includegraphics{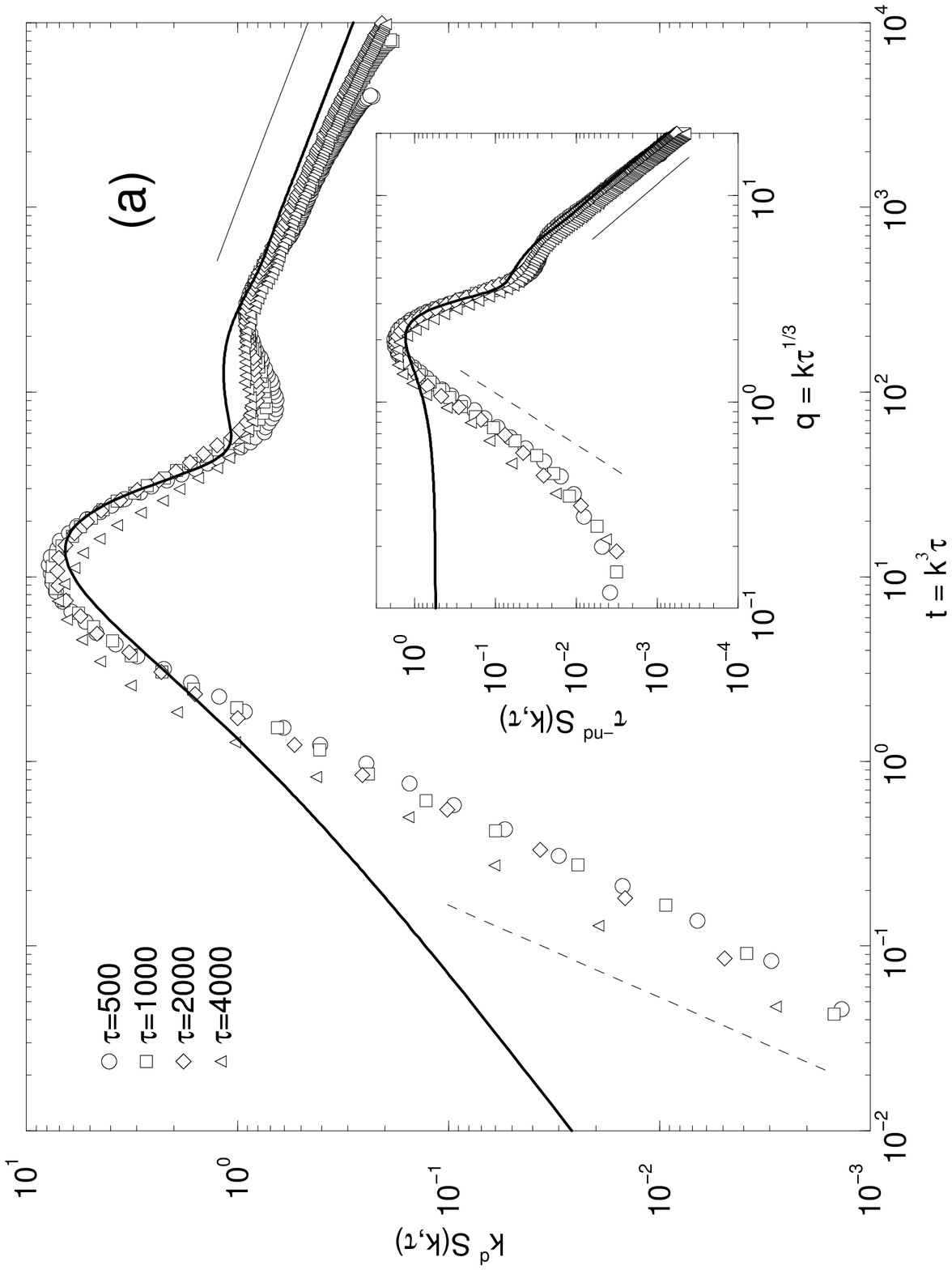}
\vskip 0.5in
\vskip 2.85in
\includegraphics{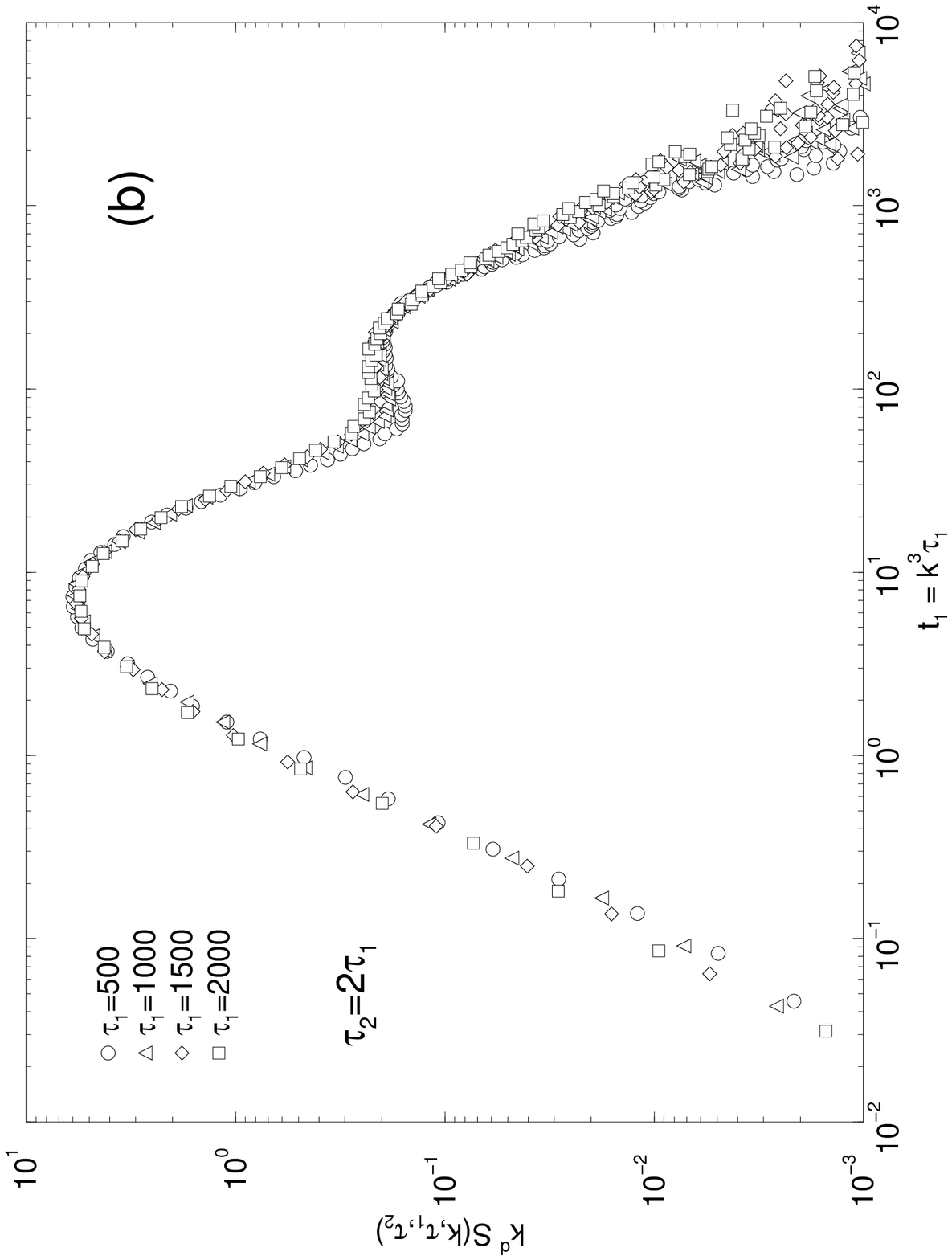}
\bigskip
\caption[]{Scaling collapse of the structure factor. (a) Scaling of the
one-time structure factor for several different times presented on a
log-log scale vs $t=k^3\tau.$ The slopes of the 
straight lines
correspond to the expected power laws, $S(k,\tau) \propto k^4$ at
small wave vectors
(dashed line)
and $ \propto k^{-(d+1)}$ at large wave vectors
(solid line),
respectively. The heavy solid curve is the theory of Yeung, Oono, and
Shinozaki [39]. The inset shows the equivalent scaling collapse of
Eq.~(3). (b) Scaling of the two-time structure factor for fixed
$\tau_1/\tau_2=1/2$ and several different $\tau_1$. This corresponds
to a cut through ${\rm Cov}(k,\tau_1,\tau_2)$ along a line of slope
$1/2$. The $\tau_1=2000$ data represented by squares are the same as
those represented by squares in Fig.~6(b).}
\label{fig:skscale}
\end{figure}

Measurements that include intensities only for two measurement times,
$\tau_1$ and $\tau_2$, correspond to rays of slope $\tau_1/\tau_2$ in
the $(t_1,t_2)$ plane. The wave vector then serves as the parameter
indicating position along the ray, and the values of the intensity
covariance along the $t_1 = t_2$ ray are the structure factor
squared. Figure~\ref{fig:skscale}(a) shows the simulation estimates
for the scaled structure factor $F_1(t)$ for several different times
on a log-log scale. The collapse for different $\tau$ is quite
reasonable. The two straight lines are provided as a reference to the
expected asymptotic behavior.  The dashed line in
Fig.~\ref{fig:skscale}(a) indicates the slope corresponding to the
power law $S \propto k^{\beta(n)},$ or $F_1 \propto
t\,^{n(\beta(n)+d)},$ with $\beta = 4$ for model~B as discussed in
Sec.~\ref{sec:smallt}.  The solid line corresponds to the Porod's law
result $S \propto k^{-(d+1)}$ (or $F_1 \propto t^{-n}$) at high wave
vectors, which is an important part of the moving-interface model
associated with the asymptotic scaling form $\Phi_d(z)$. The simulated
structure factor reproduces both behaviors. The inset is the
representation of the master curve $F'_1$ in terms of the scaled wave
vector $q = k [ B \tau ]^{1/3}.$ The deviation from the expected power
law at small $q$ has been observed in other simulations and is
expected to vanish as $\tau\rightarrow\infty$ \cite{YEUN88}.

In the case of nonconserved order parameter, the analytic theory of
Ohta, Jasnow, and Kawasaki (OJK) \cite{OHTA82,OHTA84} gives excellent
quantitative agreement for the one-time structure factor
\cite{BROW98A,NOTE98A}. The YJ extension of that theory \cite{YEUN90}
to two-time correlations gives good agreement for the intensity
covariance except at very large $\delta t$ \cite{BROW97A}.  No
comparably successful theory exists for the conserved order-parameter
case considered here. Approximate forms for the one-time structure
factor (none of which has the expected $k^4$ behavior at small $k$)
have been proposed by Ohta and Nozaki \cite{OHTA89}, and by Tomita
\cite{TOMI93}, Yeung, Oono, and Shinozaki \cite{YEUN94}. Of these, we
find the best agreement for the latter result. The heavy solid curve
in Fig.~\ref{fig:skscale}(a) represents this approximate structure
factor as obtained from the differential-equation formulation in
$k$-space found in Ref.~\cite{YEUN94}. This theory extends a model of
interface motion to phase separation and recovers $n=1/3$. Here it has
been fit numerically to the simulation data using one adjustable
parameter. While it gives reasonable agreement at small distances, it
fails at larger distances as seen by the deviations at small $t$ in
Fig.~\ref{fig:skscale}(a). There is also noticeable disagreement
around the shoulder at high $t$. A better analytic model for the
locally-conserved dynamics of phase separation is clearly still
needed.

The scaling of the two-time structure factor along the ray
$\tau_1/\tau_2=1/2$ is shown in Fig.~\ref{fig:skscale}(b) for several
$\tau_1$. The data collapse is again quite good. The largest change
from the one-time structure factor shown in Fig.~\ref{fig:skscale}(a)
is the disappearance of the Porod tail at large wave vector, which is
consistent with the idea that moving interfaces are controlling the
two-time properties of the system.

\begin{figure}[tb]
\vskip 2.85in
\includegraphics{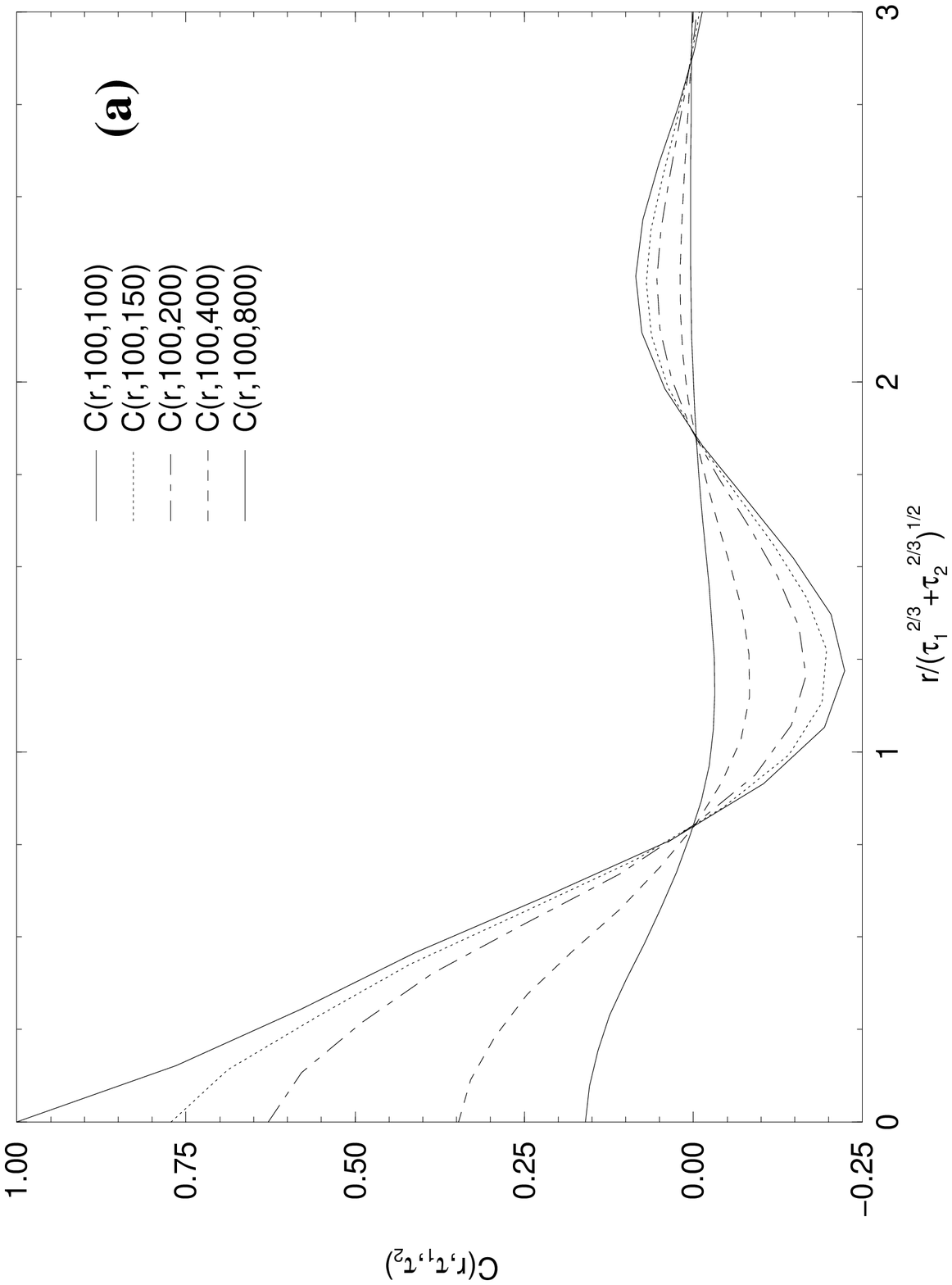}
\vskip 0.5in
\vskip 2.85in
\includegraphics{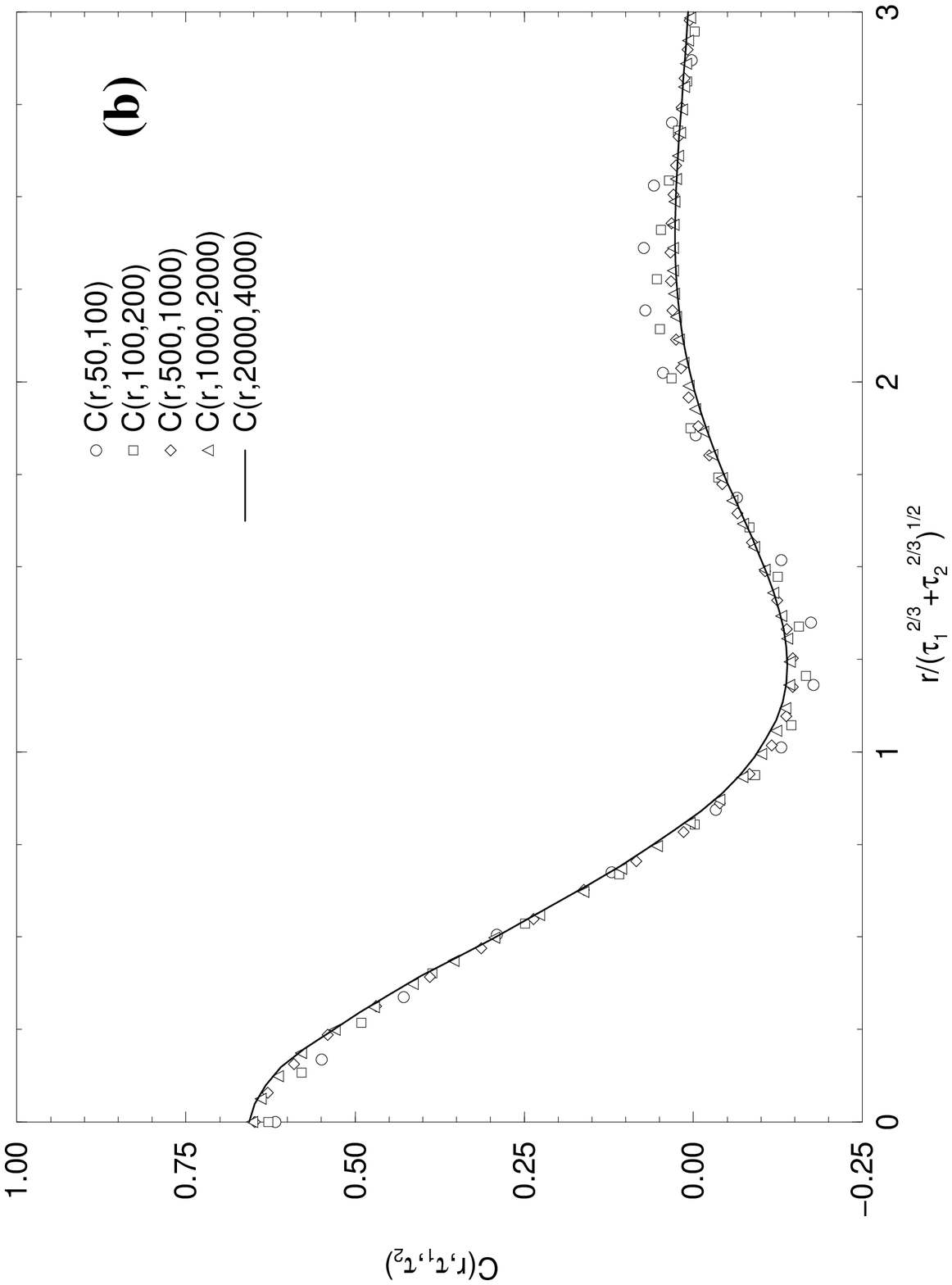}
\bigskip
\caption[]{Scaling of the two-time correlation function. (a) For
$\tau_1=100$ and various $\tau_2$. The zero crossing points remain
approximately stationary while the amplitude of oscillations decreases
monotonically. (b) For several $\tau_1/\tau_2=1/2$. A scaling form
appears to be approached at later times. The solid curve, the
simulation result involving the latest time, is taken as an estimate
of the scaling function for this ratio of times.}
\label{fig:corrscale}
\end{figure}

It is also interesting to consider scaling for the two-time
order-parameter correlation function for large $r$, where it is not
linear in $r$ and the scattering is not governed by Porod's law.  In
Sec.~\ref{sec:smallt}, we argued that the scaling variables should be
$r/[B\bar{\tau}\,]^n$ and $\delta \tau/\bar{\tau},$ and this gives
quite good agreement with the simulations. However, an equivalent set
of scaled variables gives what we believe is a more natural form for
the correlation function. It is inspired by the YJ correlation
function appropriate to model~A \cite{BROW97A,YEUN90},
Eq.~(\ref{eq:CYJ}), where scaled lengths enter as
$r^2/[R^2(\tau_1)+R^2(\tau_2)]$. (This is a consequence of the
assumption of a Gaussian auxiliary field.) The simulated two-time
order-parameter correlation function for model~B is well scaled out to
quite large $r$ using
\begin{equation}
C({\bf r},\tau_1,\tau_2)
= \tilde{C} \left(\frac{r}{B^n \sqrt{\tau_1^{2n}+\tau_2^{2n}}},
\frac{\tau_1}{\tau_2}\right) 
\;.
\label{eq:c2scale}
\end{equation}
The results for fixed $\tau_1$ and several $\tau_2$ are presented in
Fig.~\ref{fig:corrscale}(a). The zeros of the correlation function are
approximately stationary with respect to
$x=r/\sqrt{\tau_1^{2/3}+\tau_2^{2/3}}$ for the range of times shown
here. The scaling becomes less good for much smaller values of
$\tau_1/\tau_2$. As the ratio $\tau_1/\tau_2$ decreases the amplitude
of the oscillations also decreases. The collapse of
$C(r,\tau_1,\tau_2)$ onto $\tilde{C}(x,\tau_1/\tau_2)$ for several
pairs of $\tau_1$ and $\tau_2$ is shown in
Fig.~\ref{fig:corrscale}(b). The simulation results appear to converge
for later times, and the collapse is quite good for $\tau_1 \ge
500$. The solid curve represents the simulation estimate of the master
curve for this ratio of times with $\tau_2=4000$.

\begin{figure}[tb]
\vskip 2.85in
\includegraphics{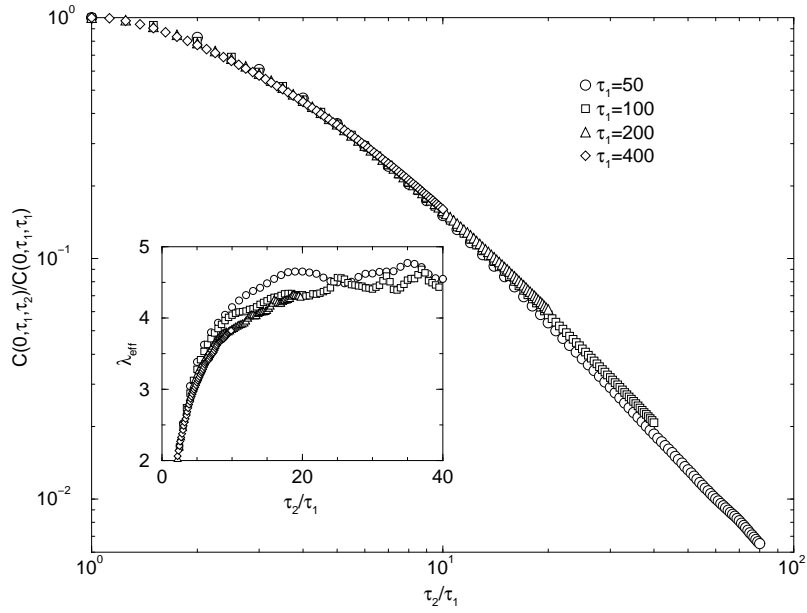}
\bigskip
\smallskip
\caption[]{Log-log plot of the autocorrelation vs $\tau_2/\tau_1$ for
several values of $\tau_1$. Least-squares fitting for $\tau_1=50$,
$\tau_2>2000$ gives a slope of $1.49 \pm 0.01$ which corresponds to
$\lambda \approx 4.5$. This value violates the upper bound $\lambda
\le d$ conjectured by Fisher and Huse [40], but is consistent with the
constraint $\lambda\ge(d/2)+2$ derived by Yeung, {\em et al.\/} [41], for
conserved systems.}
\label{fig:auto}
\end{figure}

One last property of phase-separating materials to consider is the
autocorrelation of the order parameter, $C(0,\tau_1,\tau_2)$, for
$\tau_2 \gg \tau_1$.  In this regime (the opposite of the small
time-difference regime emphasized elsewhere in this paper) Fisher and
Huse \cite{FISH88} have argued that the autocorrelation should be
described by a power-law exponent $\lambda$ as
\begin{equation}
\label{eq:autocorr}
C(0,\tau_1,\tau_2) \propto
\left(\frac{R(\tau_1)}{R(\tau_2)}\right)^\lambda \;.
\end{equation}
The YJ correlation function for model~A (see Appendix) predicts
$\lambda=d/2$. This is a weakness of the YJ theory; the value observed
in experiments and simulations on model~A indicate $\lambda$ is
clearly larger \cite{BROW97A}. For model~B, Yeung, Rao, and Desai
(YRD) \cite{YEUN96} found $\lambda \approx 4$ for $\tau_1$ in the
scaling regime from a numerical simulation similar to the one
presented here. The autocorrelation for the present simulations is
shown in Fig.~\ref{fig:auto}, where the results have been normalized
by $C(0,\tau_1,\tau_1)$ to show the collapse for different
$\tau_1$. The inset shows the values of $\lambda$ estimated from the
derivative of $C(0,\tau_1,\tau_2)$ as a function of
$\tau_2/\tau_1$. For the two earliest $\tau_1$, $\lambda$ appears to
have converged to its asymptotic value. A least-squares fit to the
latest times ($\tau_2>2000$) for $\tau_1=50$ gives $\lambda=4.47 \pm
0.03$. Both our result and that of YRD violate the upper bound
$\lambda \le d$ conjectured by Fisher and Huse for phase-ordering
systems \cite{FISH88}. They are consistent with the lower bound
$\lambda\ge (d/2)+2$ found for spinodal decomposition in $d \ge 2$ by
YRD \cite{YEUN96}. Lee and Rutenberg \cite{LEE97} found the same bound
for small, but nonvanishing, minority-phase volumes. They point out
that correlations in the minority domains accelerate the decay of the
autocorrelation.

\section{Conclusions}
\label{sec:conclusion}

The scattering intensity for two-dimensional systems undergoing
spinodal decomposition has been investigated using numerical
integration of the Cahn-Hilliard-Cook equation.  The results indicate
that the two-time structure factor can be measured experimentally via
the speckle-intensity covariance at ${\bf k} \neq 0$.  The connection
between the two quantities is established directly by comparing
numerical estimates of both and indirectly by verifying the
exponential probability distribution of the scattering intensity.  For
the present simulations, the equality between the two-time structure
factor and the speckle-intensity covariance is observed to break down
at $z \agt 4.5$. These properties were also observed in previous
simulations for non-conserved dynamics \cite{BROW97A}, and they should
be common to any heterogeneous material with a large number of
independent domains.

An important conclusion of our study is that the scaling behavior of
the speckle-intensity covariance at both large and small $\bar{t}$ can
be obtained from geometric arguments based on a simple
moving-interface picture. These scaling results depend on the dynamics
only implicitly through the values of the dynamic exponent $n$ and the
exponent $\beta(n)$, which gives the asymptotic small-$k$ behavior of
the one-time structure factor. Specifically, we have argued that the
asymptotic scaling of the two-time structure factor for large
$\bar{t}$ is in terms of $z = \delta t / \bar{t} \, ^{1-n}$, with
$1-n=2/3$ for spinodal decomposition.  For small $\bar{t},$ on the
other hand, the appropriate scaling variable is $\delta t / \bar{t} .$
These behaviors were recently confirmed by an XIFS experiment
\cite{MALI98} of spinodal decomposition and demonstrate that
experimentally observable aspects of both the short and long length-
and time-scale regimes of phase separation can be described by
relatively simple models of interface motion. A simple form motivated
by the Yeung-Jasnow theory of phase ordering gives good agreement with
simulation results for the correlation function.

Even though much of the large ${\bf k}$ behavior of model~B can be
understood theoretically, a complete theory of phase separation, even
one that just describes the one-time structure factor, is still
needed.

\section*{Acknowledgements}
We would like to thank G.~Korniss, G.~B. Stephenson, F. Bley,
F. Livet, A.~Rutenberg, and T.~Kawakatsu for useful discussions.
Research at Florida State University was supported by the Center for
Materials Research and Technology, by the Supercomputer Computations
Research Institute (under U.S.\ Department of Energy Contract No.\
DE-FC05-85ER25000), and by U.S.\ National Science Foundation grant
DMR-9634873.  Research at McGill University was supported by the
Natural Sciences and Engineering Research Council of Canada and {\em
le Fonds pour la Formation de Chercheurs et l'Aide \'a la Recherche du
Qu\'ebec.\/} Supercomputer time at the U.S.\ National Energy Research
Scientific Computing Center was made available by the U.S.\ Department
of Energy.

\section*{Appendix}

In fact, $G(x)=\sqrt{1+x^2}$ is the form which results from the YJ
theory for model~A \cite{BROW97A,YEUN90}. Those authors introduced the
two-time order-parameter correlation function \cite{YEUN90}
\begin{eqnarray}
C_{\rm YJ}(r, \tau_1, \tau_2) 
&=& \frac{2}{\pi} \arcsin 
\Bigg[ 
\left( \frac{2 R(\tau_1) R(\tau_2)}{R(\tau_1)^2 + R(\tau_2)^2} \right)^{d/2} 
\nonumber\\
& & \times \exp \left( \frac{- r^2}{R(\tau_1)^2 + R(\tau_2)^2} \right)
\Bigg] \;.
\label{eq:CYJ}
\end{eqnarray}
In this case $R(\tau) = [B \tau ]^n$ with $n = 1/2$ and $B = 4 (d-1)/d$. 
Expanding $R(\tau_1)$ and $R(\tau_2)$ in terms of $\delta \tau$ and 
$\bar{\tau}$ and expressing the argument of the $\arcsin$ to lowest order 
in $\delta \tau / \bar{\tau}$ and $r / [B \bar{\tau}]^n$ we get 
\begin{equation}
C_{\rm YJ}(r, \delta \tau, \bar{\tau}) 
= 
\frac{2}{\pi} \arcsin 
\left[ 
1 - \frac{d}{16} \left( \frac{\delta \tau}{\bar{\tau}} \right)^2 
  - \frac{1}{2} \left( \frac{r}{[B \bar{\tau}]^n} \right) + \dots 
\right]
\;.
\label{eq:CYJarg}
\end{equation}
The expansion for arguments near unity, 
$\arcsin(x) \approx \pi / 2 - \sqrt{2(1-x)}$, then gives 
\begin{equation}
C_{\rm YJ}(r, \delta \tau , \bar{\tau}) 
\approx 1 - \sqrt{\frac{d}{8}} \frac{2}{\pi} 
            \frac{n \zeta}{\bar{\tau}^{1/2}} 
\sqrt{1 + \left( \sqrt{\frac{d}{8}} \frac{r}{B^{1/2} n \zeta} \right)^2} 
\;.
\label{eq:CYJG}
\end{equation}
This correlation function corresponds to Eq.~(\ref{eq:scalfeq}) with
$C_1 = 2/ \pi$, $C_2 = \sqrt{d/8}$, and $G(x) = \sqrt{1 + x^2}$. In
Ref.~\cite{BROW97A} the resulting analytic scaling form for the
two-time structure factor at large $\bar{t}$, $\Phi_d(z)$ given by
Eq.~(\ref{eq:Phizsqrt}), was obtained for model~A by a different
method utilizing the full form, Eq.~(\ref{eq:CYJ}). That approach also
yielded a small-$\bar{t}$ result corresponding to the Taylor expansion
Eqs.~(\ref{eq:smalltexpand}) and the corresponding condition
Eq.~(\ref{eq:smalltvalid}).

~
\vfill
\newpage

~
\vfill
\newpage

~
\vfill
\newpage

~
\vfill
\newpage

~
\newpage

~
\newpage

~
\newpage

~
\newpage

~
\newpage

~
\newpage

~
\newpage

~
\newpage


\begin{thebibliography}{10}
\bibitem{k-RANGE}
The maximum dimensionless wavenumber allowed by the discretization of
the continuum equations is $\pi$.
\bibitem{NOTE98A} The result for the amplitude in Porod's law
\cite{DEBY57} was regrettably rendered incorrectly after Eq.~(9) in
Ref.~\cite{BROW98A}. Using the notation of that paper, the result
should be $F(q)=2^d \pi^{(d-1)/2} \Gamma((d+1)/2) \alpha q^{-(d+1)},$
consistent with the expressions in Eq.~(\ref{eq:tnF2z0}) of the present
paper and Eq.~(5.9) of Ref.~\cite{BROW97A}.
\end{thebibliography}
\end{document}